\documentclass{aastex}
\usepackage{emulateapj5}

\usepackage{graphicx}
\usepackage{amsmath}

\slugcomment{submitted to ApJ (2003)}

\shorttitle{A Model for Black Hole HFQPOs}
\shortauthors{Schnittman \& Bertschinger} 

\begin{document}

\title{A Model for High Frequency Quasi-periodic Oscillations from
Accreting Black Holes} 
\author{Jeremy D. Schnittman and Edmund Bertschinger}
\affil{Department of Physics, Massachusetts Institute of Technology}
\affil{77 Massachusetts Avenue, Cambridge, MA 02139}
\email{schnittm@mit.edu, edbert@mit.edu}

\begin{abstract}
Observations from the Rossi X-Ray Timing Explorer have shown
the existence of high frequency quasi-periodic oscillations (HFQPOs)
in the X-ray flux from accreting black hole binary systems. In at
least two systems, these HFQPOs come in pairs with a 2:3 frequency
commensurability. We propose a simple ``hot spot'' model to explain
the position and amplitude of the HFQPO peaks. Using the exact
geodesic equations for the Kerr metric, we calculate the trajectories of
massive test particles, which are treated as isotropic, monochromatic
emitters in their rest frames. Photons are traced from the accretion
disk to a distant observer to produce time- and frequency-dependent
images of the orbiting hot spot and background disk. The power
spectrum of the X-ray light curve consists of multiple peaks at
integral combinations of the black hole coordinate frequencies. In
particular, if the radial frequency is one-third of the azimuthal
frequency (as is the case near the inner-most stable circular orbit),
beat frequencies appear in the power spectrum at two-thirds and
four-thirds of the fundamental azimuthal orbital frequency, in
agreement with observations. We also model the effects of shearing
the hot spot in the disk, producing an arc of emission that also
follows a geodesic orbit, as well as the effects of non-planar orbits
that experience Lens-Thirring precession around the black hole
axis. By varying the arc length, we are able to
explain the different QPO features observed in ``Type A'' and
``Type B'' X-ray outbursts from XTE J1550-564. In the context of this
model, the observed power spectra allow us to infer values for the
black hole mass and angular momentum, and also constrain the
parameters of the model, such as the hot spot size and luminosity.
\end{abstract}

\keywords{black hole physics -- accretion disks -- X-rays:binaries}

\section{INTRODUCTION}
In the past decade, observations of X-ray emission from accreting
neutron stars and black holes have introduced new
possibilities for astrophysical tests of fundamental physics. Recent
discoveries made by satellites such as
ASCA, RXTE, BeppoSAX, Chandra, and XMM-Newton provide direct evidence
for strong-field gravitational effects in compact binary systems and
active galactic nuclei (AGN). These results include Doppler-broadened
iron K$\alpha$ fluorescent emission from microquasars \citep{mille02}
and millisecond variability of the X-ray flux from black 
holes in low-mass X-ray binaries
\citep{stroh01a,lamb02,remil02}. These measurements 
give the exciting prospect for determining a black hole's mass and
spin, as well as tests of general relativity in the strong-field
regime.

The strong gravitational fields near a black hole (BH) introduce
significant deviations from Newtonian physics, including the existence
of an inner-most stable circular orbit (ISCO), a feature absent in the
classical Kepler problem. Since accreting gas can efficiently lose
energy and angular momentum only outside of the ISCO, the hydrodynamic
and radiative behavior of the inner accretion disk should be strongly
dependent on the structure of the space-time metric near the
ISCO. The famous ``no hair'' theorem states that the only observable
features of an electrically neutral black hole are functions of its
mass $M$ and specific angular momentum $a\equiv J/M$. By understanding
the behavior of matter
near the ISCO, we can determine the mass and angular momentum, and
thus completely describe the black hole.

Many authors have approached the problem of accretion in compact
binaries with a variety of different methods, including early analytic
models by \citet{shaku73} and \citet{ghosh78}. 
Some have simplified the hydrodynamics in favor of a flat, thin,
steady-state disk and a more detailed treatment of general relativistic
effects \citep{laor91,karas92,reyno97}. To include dynamic effects,
essential for modeling QPOs, others have included magnetohydrodynamics
(MHD) in a pseudo-Newtonian model \citep{hawle01,armit01} 
or with smoothed particle hydrodynamics
\citep{lanza98,lee02}. A family of perturbative models has given
rise to the field of diskoseismology \citep{wagon01}, where different
global modes in the disk oscillate with different
frequencies. \citet{wang03} propose a magnetic  
coupling between the rotating black hole and the accretion disk as a
means of producing high-frequency QPOs, analogous to the
Blandford-Znajek process \citep{bland77} used to describe AGN.

Recent observations of commensurate relationships in the
high-frequency QPOs of black hole accretion disks \citep{remil02}, as
well as the longstanding puzzles of the frequency 
variability of QPO peaks and their correlations with X-ray flux and
energy \citep{lamb02}, motivate more detailed study of the QPO
phenomenon as a means to determining the black hole parameters. We
have developed a model that is a combination of many of the above
approaches in 
which additional physics ingredients can be added incrementally to a
framework grounded in general relativity. The model does not currently
include scattering, radiation pressure, magnetic fields, or
hydrodynamic forces, instead treating the
disk as a collection of cold test particles radiating isotropically
in their respective rest frames. The trajectories of emitted photons are
integrated through the metric to a distant observer to construct
time-dependent images and spectra of the disk. The dynamic model uses the
geodesic trajectory of a massive particle as a guiding center for a
small region of excess emission, a ``hot spot,'' that creates a 
time-varying X-ray signal, in addition to the steady-state background
flux from the disk.

This hot spot model is motivated by the similarity between the
QPO frequencies and the black hole (or neutron star) coordinate
frequencies near the ISCO \citep{stell98,stell99} as well as
the suggestion of a resonance leading to integer
commensurabilities between these coordinate frequencies
\citep{kluzn01,kluzn02}. \citet{stell99} investigated primarily the
QPO frequency pairs found in LMXBs with a neutron star (NS) accretor, but
their basic methods can be applied to black hole systems as well. 

\citet{marko00} have presented a thorough
analysis of this hot spot model for a collection of NS binaries for
which pairs of QPOs have been observed. Based on a number of
experimental and theoretical arguments, they conclude that the
geodesic hot
spot model is not a physically viable explanation for the observed NS 
QPOs. For low to moderate eccentricity orbits, the coordinate
frequencies simply do not agree with the QPO data. For highly
eccentric geodesics, they argue that the relative power in the
different frequency modes are qualitatively at odds with the
observations. Furthermore, they show that hydrodynamical
considerations place strong constraints on the possible size, luminosity,
coherency, and trajectories of the hot spots. 

Many of these points are addressed in our version of the hot spot
model. Also, by including full 3-dimensional (3D) ray-tracing, we can
quantitatively predict how much QPO power will be produced by a hot
spot of a given size and emissivity moving along a geodesic orbit near
the ISCO. Along with the special
relativistic beaming of the emitted radiation, we find that strong
gravitational lensing can cause high-amplitude modulations in the
light curves, even for relatively small hot spots. The issues of
differential rotation and shearing of the emission region are addressed
below when we consider the generalization of the hot spot model to
include arcs and non-planar geometries.

Perhaps the most powerful feature of this hot spot model is the
facility with which it can be developed and extended to more general
accretion disk geometries. 
In addition to providing a possible explanation for the
commensurate HFQPOs in at least two systems (XTE J1550-564 and GRO
J1655-40), the hot spot model with full general relativistic 
ray-tracing is a useful building block toward any other viable model of
a dynamic 3D accretion disk. Within the computational
framework of the Kerr metric, we can
investigate many different emission models and compare their
predicted X-ray spectra and light curves with observations. For
example, our ray-tracing code could be used in conjunction with a 3D
MHD calculation of the accretion disk to simulate the time-dependent
X-ray flux and spectrum from such a disk.

It is with this motivation that we present
the initial results of the model. Section 2 describes the relativistic
ray-tracing methods and discusses numerical techniques. In Section 3, we
develop the basic model of a steady-state disk with the potential
application to broad iron K$\alpha$ emission. Section 4 introduces the
basic hot spot model for circular geodesics and shows the effect of binary
inclination and black hole spin on the QPO power spectrum. In Section 5 the
model is extended to trajectories with non-zero eccentricity and
inclination, as well as elongated hot spots or arcs, giving a set of
model parameters that best fit the QPO data from a large number of
outbursts from XTE J1550-564. In Section 6 we present our conclusions
and a discussion of future work. 

\section{RAY-TRACING IN THE KERR METRIC}
We begin by dividing the image plane into regularly spaced ``pixels''
of equal solid angle in the observer's frame, each corresponding to a
single ray. Following the sample rays backward in time,
we calculate the original position and direction that a photon emitted
from the disk would require in order to arrive at the appropriate
position in the detector. The gravitational lensing and magnification of 
emission from the plane of the accretion disk is performed automatically
by the geodesic integration of these evenly spaced photon
trajectories, so that high magnification occurs in regions where
nearby points in the disk are projected to points with large
separation in the image plane. To model the time-varying emission from
the disk, each photon path is marked with the time delay along
the path from the observer to the emission point in the disk.

To integrate the geodesic trajectories of photons or massive
particles, we use a Hamiltonian
formalism that takes advantage of certain conserved quantities in the
dynamics. The resulting equations of motion do not contain any sign
ambiguities from turning points in the orbits, as are introduced
by many classical treatments of the geodesic equations in the Kerr
metric. We define a Hamiltonian function of eight phase space
variables $(x^\nu, p_\mu)$ and an integration variable (affine
parameter) $\lambda$ along the path length. For a general space-time
metric $g_{\mu \nu}(\mathbf{x})$ with inverse $g^{\mu
  \nu}(\mathbf{x})$, we can define a Hamiltonian $H_2$ quadratic in
the momenta as
\begin{equation}
H_2(x^\mu,p_\nu;\lambda) = \frac{1}{2}g^{\mu \nu}(\mathbf{x})p_\mu
p_\nu = -\frac{1}{2}m^2,
\end{equation}
where the rest mass $m$ is a constant ($m=0$ for photons, $m=1$ for
massive particles).

Applying Hamilton's equations from classical mechanics, we
reproduce the geodesic equations:
\begin{subequations}
\begin{equation}
\frac{dx^\mu}{d\lambda} = \frac{\partial H_2}{\partial p_\mu} = g^{\mu
\nu}p_\nu = p^\nu,
\end{equation}
\begin{equation}
\frac{dp_\mu}{d\lambda} = -\frac{\partial H_2}{\partial x^\mu} =
-\frac{1}{2}\frac{\partial g^{\alpha \beta}}{\partial x^\mu} p_\alpha
 p_\beta = g^{\gamma \beta}\Gamma^\alpha_{\mu \gamma}p_\alpha p_\beta.
\end{equation}
\end{subequations}

For any metric, the Hamiltonian $H_2$ is independent of the affine
parameter $\lambda$, allowing us to use one of the coordinates as the
integration parameter and reduce the dimensionality of the phase space
by two. We use the coordinate $t=x^0$ as the time coordinate for the
six dimensional phase space $(x^i,p_i)$. The corresponding Hamiltonian
is 
\begin{equation}
H_1(x^i,p_i;t) \equiv -p_0 = \frac{g^{0i}p_i}{g^{00}} +
\left[\frac{g^{ij}p_ip_j + m^2}{-g^{00}}
+\left(\frac{g^{0i}p_i}{g^{00}}\right)^2 \right]^{1/2}
\end{equation}
with equations of motion
\begin{subequations}
\begin{equation}\label{hameq_1}
\frac{dx^i}{dt} = \frac{\partial H_1}{\partial p_i},
\end{equation}
\begin{equation}\label{hameq_2}
\frac{dp_i}{dt} = -\frac{\partial H_1}{\partial x^i}.
\end{equation}
\end{subequations}
We have thus reduced the phase space to the six-dimensional tangent
bundle $(x^i,p_i)$. Moreover, because the metric is independent of
$t=x^0$, $H_1=-p_0$ is also an integral of motion.

In Boyer-Lindquist coordinates $(t,r,\theta,\phi)$, the Kerr metric
may be written 
\begin{equation}
ds^2 = -\alpha^2 dt^2 +\varpi^2(d\phi -\omega dt)^2 
+\frac{\rho^2}{\Delta}dr^2 +\rho^2d\theta^2,
\end{equation}
which allows a relatively simple form of the inverse metric
\begin{equation}
g^{\mu \nu} = \left(\begin{array}{cccc}
-1/\alpha^2 & 0 & 0 & -\omega/\alpha^2 \\
0 & \Delta/\rho^2 & 0 & 0 \\
0 & 0 & 1/\rho^2 & 0 \\
-\omega/\alpha^2 & 0 & 0 &
1/\varpi^2-\omega^2/\alpha^2 \end{array}\right).
\end{equation}
For a black hole of mass $M$ and specific angular momentum $a=J/M$,
we define (in geometrized units with $G=c=1$)
\begin{subequations}
\begin{eqnarray}
\rho^2 &\equiv& r^2 +a^2 \cos^2\theta \\
\Delta &\equiv& r^2 -2Mr + a^2 \\
\alpha^2 &\equiv& \frac{\rho^2 \Delta}{\rho^2 \Delta+2Mr(a^2+r^2)} \\
\omega &\equiv& \frac{2Mra}{\rho^2\Delta + 2Mr(a^2+r^2)} \\
\varpi^2 &\equiv& \left[\frac{\rho^2\Delta +
2Mr(a^2+r^2)}{\rho^2}\right] \sin^2\theta.
\end{eqnarray}
\end{subequations}
As a check, we see that the metric reduces to the well-known
Schwarzschild metric in the limit $a\to 0$. In the limit $M\to 0$,
it reduces to flat space-time with hyperbolic-elliptical coordinates.

With this form of the metric, the Hamiltonian $H_1$ can be written
\begin{equation}
H_1(r,\theta,\phi,p_r,p_\theta,p_\phi;t) = \omega p_\phi
+\alpha\left(\frac{\Delta}{\rho^2}p_r^2 
+\frac{1}{\rho^2}p_\theta^2 +\frac{1}{\varpi^2} p_\phi^2
+m^2\right)^{1/2}.
\end{equation}
This new Hamiltonian is also independent of $\phi$ (azimuthally
symmetric space-time), giving the conjugate momentum $p_\phi$ as
the second integral 
of motion for $H_1$. We are now left with five coupled equations for
$(r, \theta, \phi, p_r,p_\theta)$. The third integral of
motion, Carter's constant \citep{carte68}
\begin{equation}
\mathcal{Q} \equiv p_\theta^2
+\cos^2\theta \left[a^2(m^2-p_0^2)+p_\phi^2/\sin^2\theta\right],
\end{equation}
is used as an independent check of the accuracy of the numerical
integration. 

As mentioned above, many traditional schemes to calculate
trajectories in the Kerr metric use this additional integral of motion
to further reduce the dimensionality of the problem \citep{mtw73}
and even reduce the problem to one of quadrature
integration [e.\ g.\ \citet{rauch94}]. While potentially increasing
the speed of the computation, this approach also introduces
significant complications in the form of arbitrary signs in the
equations of motion corresponding to turning points in $r$ and
$\theta$. Another common approach is to define effective potentials in
$U_{\rm eff}(r)$ and $V_{\rm eff}(\cos\theta)$, which give a set of four
parametric equations for $r(\lambda), \theta(\lambda), \phi(\lambda),$
and $t(\lambda)$. These parametric solutions introduce the difficulty of
reconstructing the coordinate momenta, which are not evolved along 
with the coordinate positions. In our ray-tracing algorithm, we have
traded a modest improvement in computational speed for simplicity in
implementation.

The initial conditions for the photon or particle geodesics are
determined in the local orthonormal frame of a ``Zero-Angular Momentum
Observer'' (ZAMO). The ZAMO basis is defined such that the spatial
axes are aligned with the coordinate axes and then the time axis is
determined by orthogonality \citep{barde72}. The ZAMO tetrad
$\mathbf{e}_{\hat{\mu}}$ 
is 
\begin{subequations}
\begin{eqnarray}
\mathbf{e}_{\hat{t}} &=& \frac{1}{\alpha}\mathbf{e}_t +
\frac{\omega}{\alpha} \mathbf{e}_\phi \\
\mathbf{e}_{\hat{r}} &=&
\sqrt{\frac{\Delta}{\rho^2}}\, \mathbf{e}_r \\
\mathbf{e}_{\hat{\theta}} &=&
\sqrt{\frac{1}{\rho^2}}\, \mathbf{e}_\theta \\
\mathbf{e}_{\hat{\phi}} &=&
\sqrt{\frac{1}{\varpi^2}}\, \mathbf{e}_\phi,
\end{eqnarray}
\end{subequations}
where $\mathbf{e}_\mu$ is the standard coordinate basis in
Boyer-Lindquist coordinates. One advantage of the ZAMO basis is that
the basis vector $\mathbf{e}_{\hat{t}}$ is time-like
($g_{\hat{t}\hat{t}} < 0$) everywhere outside of the horizon. For a
coordinate stationary observer, on the other hand, the time basis
vector $\mathbf{e}_t$ becomes space-like ($g_{tt} > 0$) inside the
ergosphere $r_{\rm erg} = M+\sqrt{M^2-a^2\cos^2\theta}$.
For sufficiently large values of the spin parameter
$a$, the inner-most stable circular orbit (often taken for the inner
edge of the accretion disk) extends within the ergosphere, emphasizing
the advantage of using the ZAMO basis.

At a point far away from the black hole, the space-time is nearly flat so
Euclidean spherical geometry gives the spatial direction of the photon
$n^{\hat{i}}\mathbf{e}_{\hat{i}}$, from which the initial momentum in
the coordinate basis is calculated:
\begin{subequations}
\begin{eqnarray}
p_t &=& -E_{\rm obs}(\omega\varpi n^{\hat{\phi}}+\alpha) \\
p_r &=& E_{\rm obs}\sqrt{\frac{\rho^2}{\Delta}}\, n^{\hat{r}} \\
p_\theta &=& E_{\rm obs}\sqrt{\rho^2}\, n^{\hat{\theta}} \\
p_\phi &=& E_{\rm obs}\sqrt{\varpi^2}\, n^{\hat{\phi}},
\end{eqnarray}
\end{subequations}
where the photon energy measured by the distant ZAMO is $E_{\rm obs}$.

\begin{figure}[tp]
\caption{\label{plotone} Schematic picture of ray-tracing method
  from distant observer through a disk of angular thickness
  $\Delta\theta$. The rays either terminate at the black hole horizon or
  pass through the disk, with each point of intersection labeled with
  the photon position and momentum $(x^\mu,p_\mu)$.} 
\begin{center}
\includegraphics{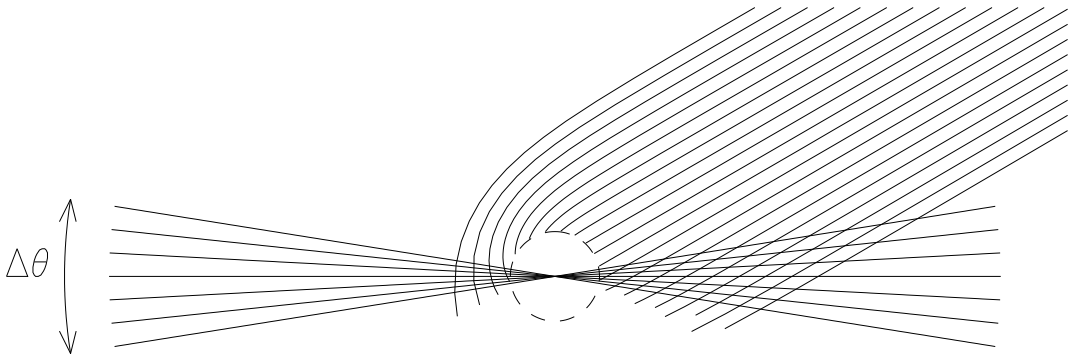}
\end{center}
\end{figure}

The photon trajectories are integrated backward in time from the
image plane oriented at some inclination angle $i$ with respect to the
axis of rotation for the black hole, where $i=0^\circ$ corresponds to
a face-on view of the disk and $i=90^\circ$ is an edge-on view. The
accretion disk is confined to 
a finite region of latitude with angular thickness $\Delta \theta$
normal to the
rotation axis. The photons terminate either at the event horizon or
pass through the surfaces of colatitude $\theta={\rm const}$, as
shown in Figure \ref{plotone}. As trajectories pass through
the disk, the photon's position and momentum $(x^\mu, p_\mu)$ are
recorded for each plane intersection in order to later reconstruct an
image of the disk.

\begin{figure}[tp]
\caption{\label{plottwo} Projection of a uniform Cartesian grid in the image
plane onto the source plane of the accretion disk $(\theta=\pi/2)$. 
Inclination angles 
are $i=0^\circ$ $(top)$ and $i=60^\circ$ $(bottom)$ and spin parameters are
$a/M=0$ $(left)$ and $a/M=0.95$ $(right)$. The region inside the horizon
is cut out from each picture.}
\begin{center}
\includegraphics{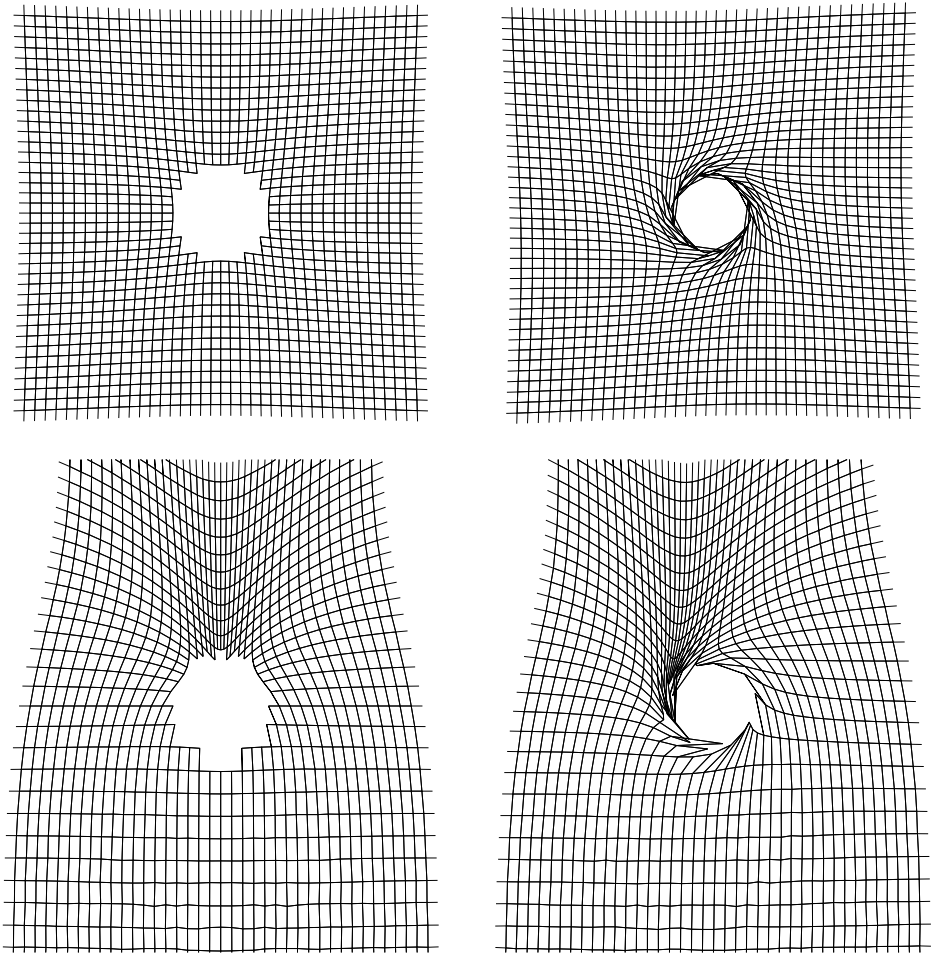}
\end{center}
\end{figure}

For an infinitely thin disk $(\Delta \theta \to 0)$, it is easy to
show how the image plane 
maps onto the source plane. Taking an evenly spaced grid of initial
photon directions, Figure \ref{plottwo} shows the positions of
intersection with the source plane, in pseudo-Cartesian coordinates
defined by 
\begin{eqnarray}\label{cart_boyer}
x &=& \sqrt{r^2+a^2}\cos \phi \nonumber\\
y &=& \sqrt{r^2+a^2}\sin \phi.
\end{eqnarray}
Photons that cross the black hole event horizon
before intersecting the plane are not shown. As the rays are deflected
by the black hole, they tend to be focused on the far side, giving a
strong magnification by mapping a large area in the image plane onto a
small area of the source plane, as seen here by a higher
density of lattice grid points. For the flat disk geometry, rays
are not allowed to pass through the plane defined by $\theta=0$, so we
do not see multiple images of sources ``behind'' the black hole, as is
often observed in the strong gravitational lensing of distant quasars by
intervening galaxies. However, for sufficiently high inclinations and
spin values,
single points in the equatorial plane can be mapped to different regions of
the image plane, creating multiple images of certain regions of the
disk. This effect is seen in the folding of the image map onto itself
near the horizon in the bottom right of Figure \ref{plottwo}.

The disk itself is modeled as a collection of mass 
elements moving along circular geodesics around the black hole, emitting
isotropic, monochromatic light with energy $E_{\rm em}$ in the emitter's
rest frame. For each photon with 4-momentum $p_\mu(\mathbf{x}_{\rm
  em})$ intersecting a particle trajectory with coordinate velocity 
$v^\mu(\mathbf{x}_{\rm em})$, the measured redshift is given by
\begin{equation}
\frac{E_{\rm obs}}{E_{\rm em}} =
\frac{p_\mu(\mathbf{x}_{\rm obs})v^\mu(\mathbf{x}_{\rm obs})}
{p_\mu(\mathbf{x}_{\rm em})v^\mu(\mathbf{x}_{\rm em})},
\end{equation}
where for a distant observer at $r\to \infty$, we take
$v^\mu(\mathbf{x}_{\rm obs}) =
[1,0,0,0]$. 

For disk models with finite thickness, the radiation transport
equation can be solved as the ray passes through the disk. In
a locally flat frame, the radiation transport equation is
\begin{equation}
\frac{dI_\nu}{ds} = j_\nu -\kappa_\nu I_\nu,
\end{equation}
where $ds$ is the differential path length and $I_\nu$, $j_\nu$, and
$\kappa_\nu$ are respectively the radiation intensity and the emissivity
and opacity of the plasma at a frequency 
$\nu$ (scattering terms can be included in both $j_\nu$ and
$\kappa_\nu$). Additionally, as the photon bundle propagates through
the global curvature around the black hole, the spectral intensity at
a given frequency evolves as the photons are gravitationally
red-shifted, maintaining the Lorentz invariance of $I_\nu/\nu^3$.

For most of the calculations presented in this paper, we are primarily
concerned with radiation coming from a limited region of the disk, treated as
a monochromatic source with $\kappa_\nu = 0$ and $j_\nu(\mathbf{x}) =
\delta(\nu-\nu_{em})g(\mathbf{x})$, with $g(\mathbf{x})$ the 
emissivity at space-time coordinate $\mathbf{x}$. When calculating the
emission from a flat, steady-state disk, the plane defined by $\cos
\theta=0$ is taken to be totally opaque $[g(\mathbf{x}) \propto
  \delta(\cos\theta); \kappa_\nu(\cos\theta < 0) \to \infty]$ so that
rays cannot curve around and see the ``underside'' of the accretion disk.
For each pixel $(i,j)$ in the image plane, an observed photon
bundle spectrum $I_\nu(t_{obs},i,j)$ is given for each time
step by integrating the contribution of the hot spot and the disk
through the computational grid. This collection of incident photons
can then be summed to give time-dependent light curves, spectra, or spatially
resolved images. The results of this paper are based primarily on
flat, steady state disks and single hot spots, yet with the
computational methods described above, arbitrary disk geometries and
emissivity/opacity models can be simulated as well. 

The ray-tracing calculation is carried out by numerically integrating
equations (\ref{hameq_1}) and (\ref{hameq_2}) with a fifth-order
Runge-Kutta algorithm with adaptive time stepping. This provides high
accuracy over a large range of scales as the photon follows a long
path through the relatively flat space-time between the observer and the
black hole, and then experiences strong curvature over a small region
close to the horizon. We typically maintain an accuracy of one part in
$10^{11}$, which can be independently confirmed by monitoring
$\mathcal{Q}$, Carter's constant. The images and spectra are formed by
ray-tracing a set of photon paths, usually of the order $400\times
400$ grid points in $(i,j)$ with $\sim 20$ latitudinal zones in
$\theta$ and spectral resolution of $\nu/\Delta \nu \sim 200$.

\section{SPECTRA FROM STEADY-STATE DISKS}
A steady-state disk can be made of a collection of massive particles 
moving in concentric planar circular orbits. For orbits at a radius
$r$ in a plane orthogonal to the spin axis, a particle's specific
energy and angular momentum are given analytically by \citet{shapi83}:
\begin{equation}
-p_0 = \frac{r^2-2Mr\pm a\sqrt{Mr}}{r(r^2-3Mr\pm 2a\sqrt{Mr})^{1/2}}
\end{equation}
and
\begin{equation}
p_\phi = \pm\frac{\sqrt{Mr}(r^2\mp 2a\sqrt{Mr}+a^2)} {r(r^2-3Mr\pm
 2a\sqrt{Mr})^{1/2}}.
\end{equation}
Here the top sign is taken for prograde orbits (particle angular
momentum parallel to black hole angular momentum) and the bottom sign
for retrograde orbits. Inside the ISCO, the particles follow plunge
trajectories with constant energy and angular momentum determined at
the ISCO. In practice, when calculating emission from a steady-state
disk, we take the inner edge of the disk to be the ISCO radius $R_{\rm
  ISCO}$.

For a disk made up of massive particles on circular orbits emitting
isotropically from a 
region between $R_{\rm in}$ and $R_{\rm out}$, the Doppler broadening of an
emission line (typically iron K$\alpha$ with $E_{\rm em} \approx 6.5$
keV) may be used to determine 
the inclination of the disk with respect to the
observer. Disks at higher inclination will have an intense blue-shifted
segment of the spectrum corresponding to the Doppler-boosted photons
emitted from gas moving toward the observer. The higher intensity for
the blue-shifted photons is caused by relativistic beaming, determined
by the Lorentz invariance of $I_\nu/\nu^3$ along a photon bundle:
\begin{equation}
I_\nu(\rm obs) = I_\nu(\rm em) \frac{\nu^3_{\rm obs}}{\nu^3_{\rm em}}.
\end{equation}

\begin{figure}[tp]
\caption{\label{plotthree} Normalized spectra of a monochromatic
  emission line from steady-state
  accretion disks of varying inclination. An inclination of $i=0^\circ$
  corresponds to a face-on view of the disk while $i=90^\circ$ would be
  edge-on. The emissivity is taken to be uniform between $R_{\rm in} =
  R_{\rm ISCO}$ and $R_{\rm out} = 15M$. The spin is taken to be
  $a/M=0.5$ but the dependence on $a$ is negligible for
  uniformly emitting disks.} 
\begin{center}
\includegraphics{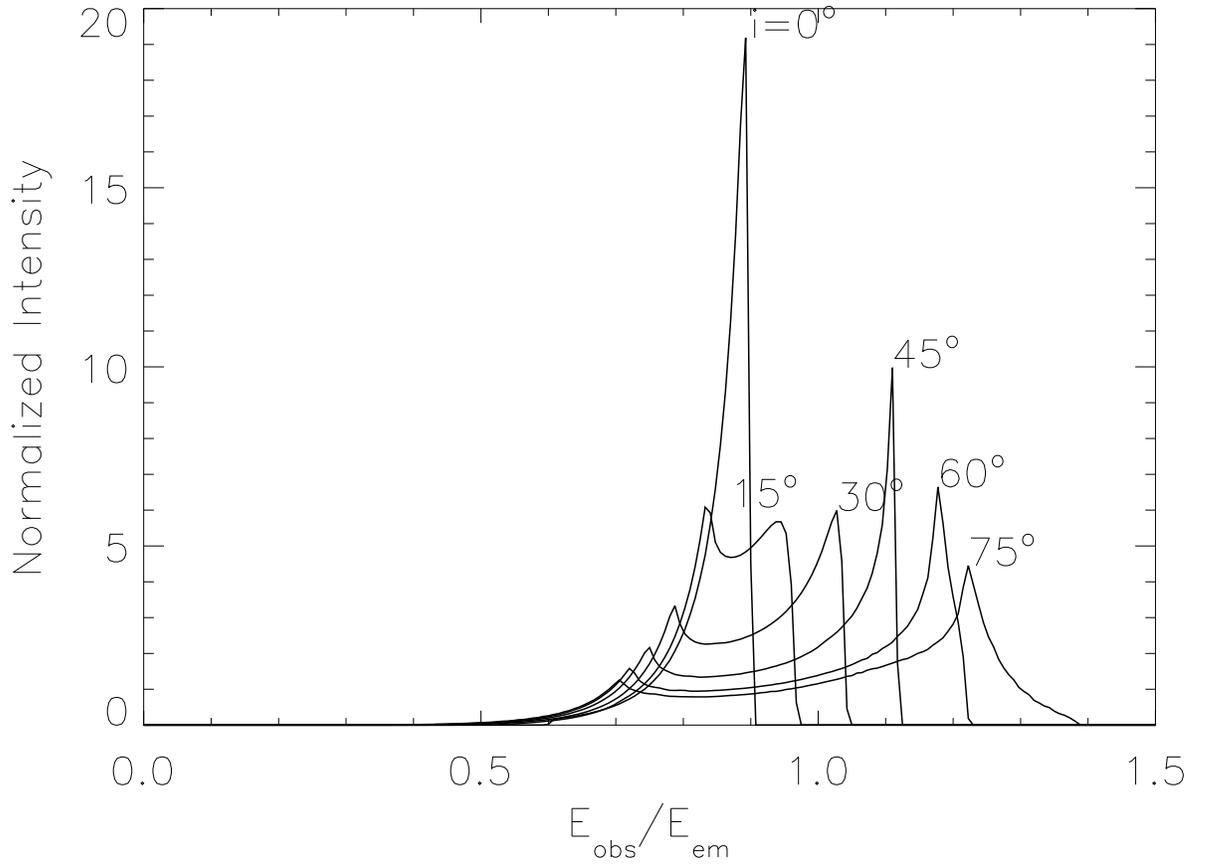}
\end{center}
\end{figure}

Figure \ref{plotthree} shows the integrated spectra from a set of
accretion disks with outer radius $R_{\rm out}=15M$ and inner radius at
$R_{\rm in}=R_{\rm ISCO}$ for a spin parameter $a/M = 0.5$, normalized
such that $\int I(E_{\rm obs}/E_{\rm em}) d(E_{\rm obs}/E_{\rm em}) =
1$. All spectra are 
assumed to come from a flat, opaque disk with uniform emission
$[g(\mathbf{x})\propto \delta(\cos\theta)]$. Repeating this
calculation for a range of spin parameters (and thus a range of
$R_{\rm ISCO}$), we find that the dependence on disk inclination is
quite strong,
while the dependence on black hole spin is almost insignificant. This
is reasonable because, except for very close to the horizon, the spin has
little effect on the orbital velocity for circular orbits: for spin
parameter $a$ and radius $r$, the observed angular frequency is given by
\begin{equation}\label{omega_phi}
\Omega_\phi = \frac{\pm \sqrt{M}}{r^{3/2}\pm a\sqrt{M}},
\end{equation}
where again we take the upper sign for prograde orbits and the lower
sign for retrograde orbits. For the large part of the disk, the orbits
have nearly Keplerian 
orbital frequencies, as measured in coordinate time $t$. For larger
values of $a$, the ISCO extends in close to  the event horizon,
increasing the radiative area of the disk. However, due to
the strong gravitational redshift in this inner region, the observed
intensity is reduced by a significant factor of $\nu^3_{\rm
  obs}/\nu^3_{\rm em}$, resulting in a weak dependence on spin for disks
with uniform emission.

Many accretion disk models include an emissivity that scales as a
power of the radius. Following \citet{broml97}, we apply an emissivity
factor proportional to $r^{-2}$, giving an added weight to the
inner, presumably hotter, regions. For the inner section of an
accretion disk with $R_{\rm out}=15M$, this inverse-square dependence
closely approximates the classic Shakura-Sunyaev steady-state disk model
\citep{shaku73}. 

As can be seen in Figure 4, this extra emission from close to the
black hole serves to break the otherwise weak dependence on spin.
For an inclination of $i=30^\circ$, five different spin values
are shown: $(a/M=-0.99, -0.5, 0, 0.5, 0.99)$, corresponding to inner disk
boundaries at $(R_{\rm ISCO}/M = 8.97, 7.55, 6.0, 4.23, 1.45)$. Since the sign
of $a$ is defined with respect to the angular momentum of the
accretion disk, negative values of $a$ imply retrograde orbits that do
not survive as close to the black hole, plunging at larger values of
$R_{\rm ISCO}$. The disks that extend in closer produce more
low-energy red-shifted
photons, giving longer tails to the spectra at $E_{\rm obs}/E_{\rm em}
< 0.7$ and smaller relative peaks at $E_{\rm obs}/E_{\rm em} \approx 1$. 

Thus, if we can determine
the inclination of the disk independently (e.g.\ through spectroscopic
observations of the binary companion), the spin may be
inferred from the broadening of an iron emission line. However, since
the plane of the disk tends to align normal to the black hole spin
axis close to the ISCO, the binary inclination may not coincide
with the inclination of the inner disk. The problem of inclination,
along with other complications, such as
additional emission lines and other causes of scattering and line
broadening, motivates us to look more closely at the QPO spectra as a
method for determining black hole spin.

\section{HOT SPOT EMISSION}
Given the map from the accretion disk to the image plane, with each
photon bundle labeled with a distinct 4-momentum and time delay, we
can reconstruct time-dependent images of the disk based on time-varying
emission models. The simplest model we consider is a single region
of isotropic,
monochromatic emission following a geodesic trajectory:
the ``hot spot'' or ``blob'' model \citep{stell98,stell99}. 

The hot spot is a small region with finite radius and
emissivity chosen to have a Gaussian distribution in local Cartesian space:
\begin{equation}\label{I_exp}
g(\mathbf{x}) \propto \exp\left[-\frac{|\mathbf{\vec{x}}-
\mathbf{\vec{x}}_{\rm spot}(t)|^2}{2 R_{\rm spot}^2}\right]. 
\end{equation}
The spatial position 3-vector $\mathbf{\vec{x}}$ is given in pseudo-Cartesian
coordinates by the transformation defined by equation (\ref{cart_boyer})
and $z=r\cos\theta$. Outside a distance of $4R_{\rm spot}$ from the
guiding geodesic trajectory, there is no emission. We typically take
$R_{\rm spot}=0.25-0.5M$, but find the normalized light curves and QPO power
spectra to be independent of spot size. Because we assume all points
in the hot spot have the same 4-velocity as the geodesic guiding
trajectory, one must be careful not to use too large a spot or the
point of emission $\mathbf{x}$ can be spatially far enough away from
the center $\mathbf{x}_{\rm spot}$ to render the inner product
$p_\mu(\mathbf{x})v^\mu(\mathbf{x}_{\rm spot})$ unphysical. We have
also explored a few different hot spot shapes, ranging from spherical
to a disk flattened in the $\hat{\theta}$ direction and similarly find
no significant dependence of the spectra on spot shape.

After calculating and tabulating the hot spot trajectory as a function
of coordinate time $t$, the ray-tracing map between the disk and the
observer is used to construct a time-dependent light curve from the
emission region. For each photon bundle intersection point there is a
time delay $\Delta t_{i,j,k}$ (where $i,j$ are the coordinate indices
in the image plane and $k$ is the latitude index in the disk) so for
the observer time $t_{\rm obs}$, we first determine where the hot spot
was at coordinate time $(t_{\rm em})_{i,j,k} =
t_{\rm obs}-\Delta t_{i,j,k}$. If the spot centroid is close (within
$4R_{\rm spot}$) to the disk intersection point $(r,\theta,\phi)_{i,j,k}$,
then the redshifted emission is added to the pixel spectrum
$I_\nu(t_{\rm obs},i,j)$, weighted by equation (\ref{I_exp}). 

In this way, a movie can be produced that shows
the blob orbiting the black hole, including all relativistic
effects. Such a movie shows a few
immediately apparent special relativistic effects such as the Doppler
shift and beaming as the spot moves toward and then away from the
observer. For a hot spot orbiting in the clock-wise direction as seen
from above ($v^\phi <0$ with $\phi=270^\circ$ toward the observer),
the point of maximum blue
shift actually occurs at a point where $\phi > 0$ because of the
gravitational lensing of the light, beamed in the forward direction of
the emitter and then bent toward the observer by the black
hole. Gravitational lensing also causes significant magnification
of the emission region when it is on the far side of the black hole,
spreading the image into an arc, much like distant galaxies are distorted
by intervening matter in galaxy clusters.

A simulated time-dependent spectrum or \textit{spectrogram} for this
hot spot model is shown 
in Figure \ref{plotfive}. The  horizontal axis measures time in the
observer's frame, with $t=0$ corresponding to the time at which the
spot center is moving most directly away from the observer $(\phi =
180^\circ)$. As mentioned above, this is \textit{not} the same as the
point of maximum redshift, which occurs closer to $\phi = 160^\circ$ due
to gravitational deflection of the emitted light.

\begin{figure}[tp]
\caption{\label{plotfour} Normalized spectra of steady-state
  accretion disks with inclination $a=30^\circ$ and varying spin
  parameter $a$ (negative values of $a$ correspond to retrograde disk
  rotation). The emissivity is taken to be proportional to $r^{-2}$ 
  between $R_{\rm in} = R_{\rm ISCO}$ and $R_{\rm out} = 15M$. Black
  holes with
  higher values of $a$ allow the inner disk to extend in closer to the
  horizon, giving a greater weight to the high-redshift radiation
  emitted there.} 
\begin{center}
\includegraphics{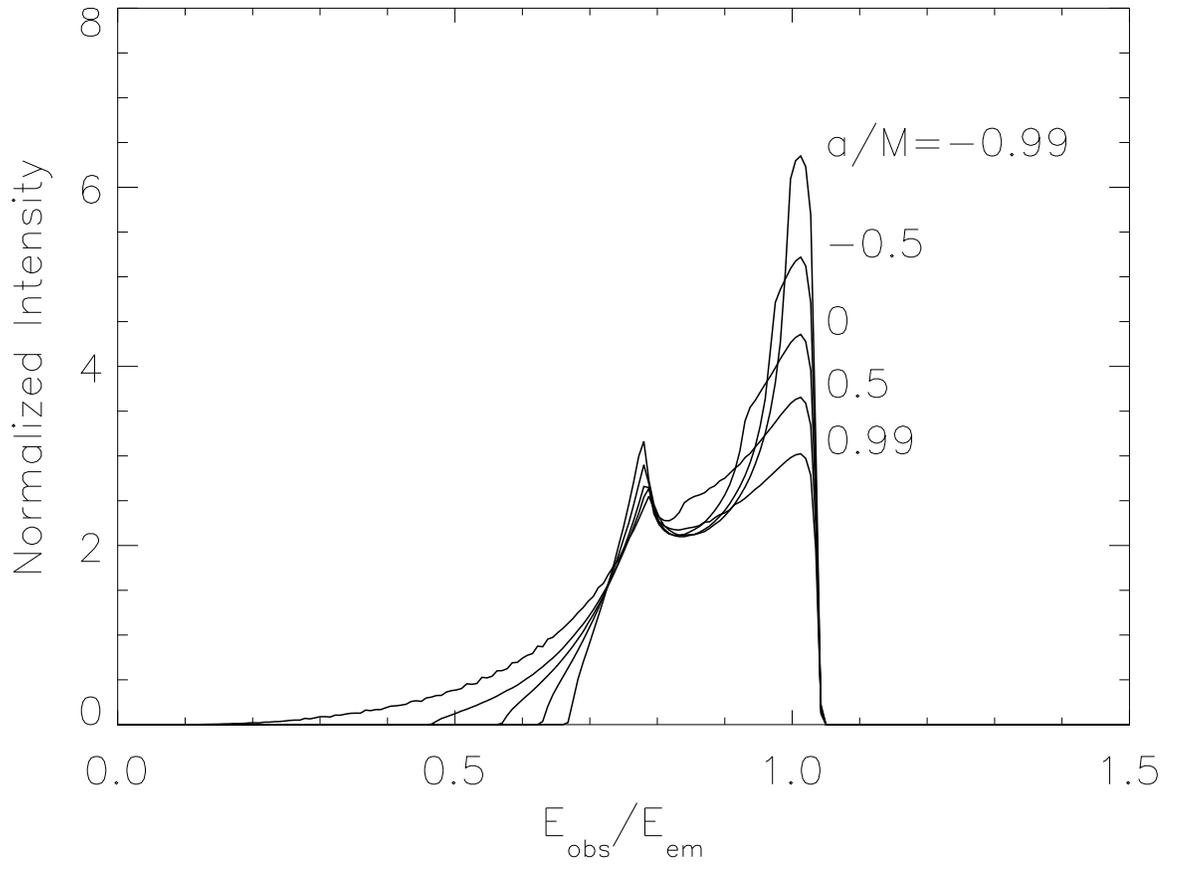}
\end{center}
\end{figure}

\begin{figure}[tp]
\caption{\label{plotfive} Spectrogram of a circular hot spot with
  radius $R_{\rm spot} = 0.5M$ orbiting a
  Schwarzschild black hole at the ISCO $(R_{\rm ISCO}=6M$), viewed at an
  inclination of $60^\circ$. The spot is moving in the
  $-\mathbf{e}_{\hat{\phi}}$ direction with the observer at
  $\phi=270^\circ$. The maximum
  redshift occurs when $\phi \approx 160^\circ$ and the maximum blueshift
  occurs when $\phi \approx 20^\circ$.}
\begin{center}
\includegraphics{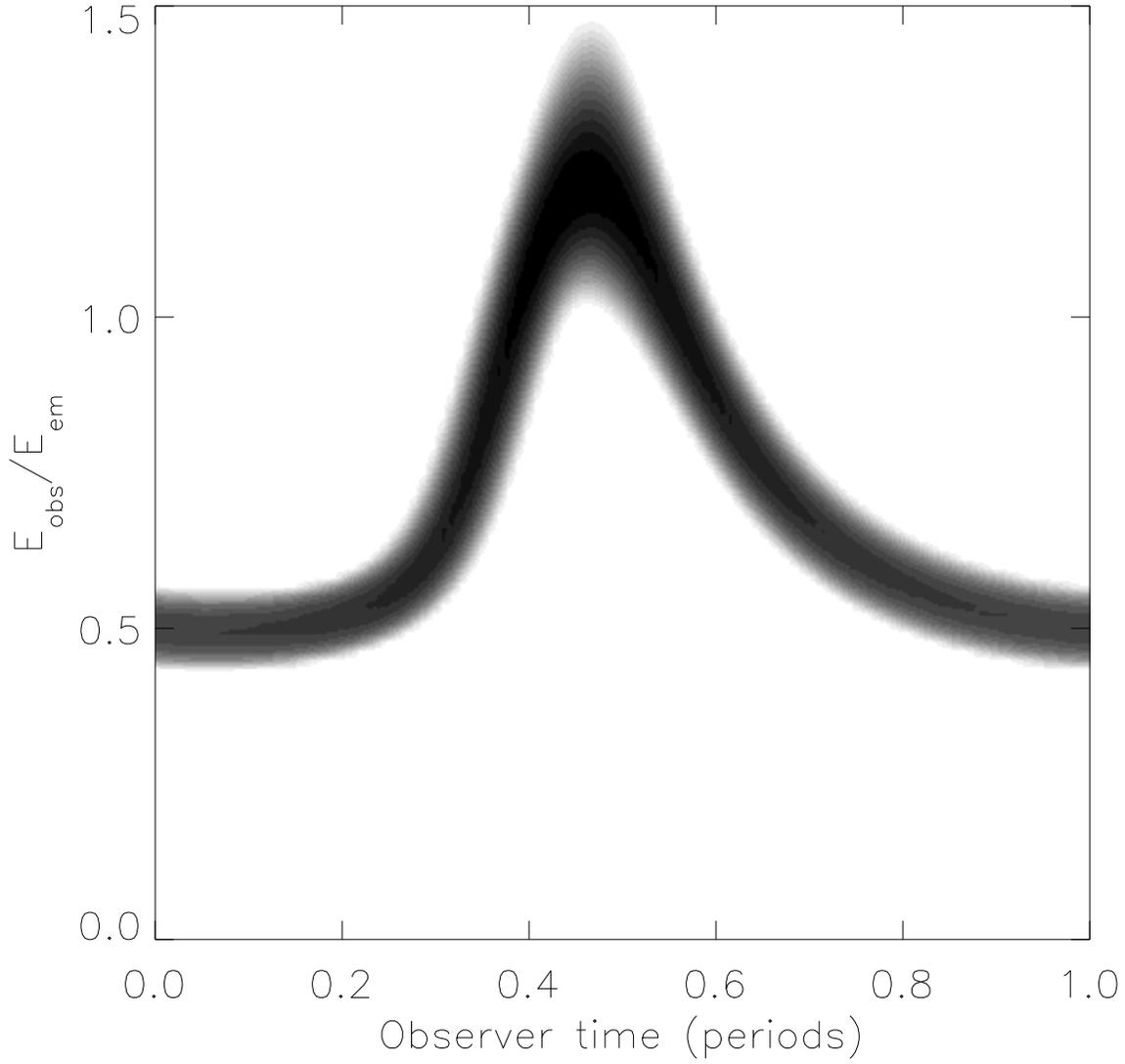}
\end{center}
\end{figure}

The spectrum shown in Figure \ref{plotfive} can be integrated in time
to give a spectrum similar to those shown in Figures
\ref{plotthree} and \ref{plotfour},
corresponding to something like a very narrow circular emitting region with
$R_{\rm out} \approx R_{\rm in} \approx R_{\rm ISCO}$. By integrating
over frequency, we get the total X-ray flux as a 
function of time, i.e.\ the light curve $I(t)$. This time-varying signal
can be added to a background intensity coming from the inner regions
of a steady state disk described in Section 3. By definition the hot
spot will have a higher temperature or density and thus greater
emissivity than the background disk, adding a small modulation to the
total flux. RXTE observations find the HFQPO X-ray modulations to have
typical amplitudes of 1-5\% of the mean flux during the outburst
\citep{remil02}. \citet{marko00} present a
first-order argument that a 1\% amplitude modulation requires a hot
spot with 100\% overbrightness extending over an area of 1\% of the
steady-state region of the disk. For $R_{\rm out}=15M$, this requires a
hot spot with radius $R_{\rm spot} \approx 1.5M$, which they argue is too
large to survive the viscous shearing of the disk.

Hydrodynamic stability aside, this reasoning ignores the
important effects of disk inclination, relativistic beaming and
gravitational lensing of the hot spot emission. Also, assuming a
Shakura-Sunyaev type disk with steady-state emissivity $g(r) \propto
r^{-2}$ and a similar scaling for the hot spot emission,
we find that hot spots with smaller size and overbrightness are
capable of creating X-ray modulations on the order of 1\% rms, defined
by
\begin{equation}\label{rms}
{\rm rms} \equiv \sqrt{\frac{\int\left[I(t)-\bar{I}\right]^2dt}
  {\int I(t)dt}}.
\end{equation} 

\begin{figure}[tp]
\caption{\label{plotsix} Overbrightness required of a hot spot on a
  circular orbit to produce a 1\% rms modulation in X-ray flux when
  added to a steady-state
  disk with $R_{\rm in}=R_{\rm ISCO}$, $R_{\rm out}=15M$ and emissivity
  $g(r)\propto r^{-2}$. An overbrightness of unity means the peak hot
  spot emissivity is twice that of the steady-state disk with no hot
  spot. The spot size $R_{\rm spot}$ is measured in gravitational
  radii $M$, so for a black hole with $a/M=0.5$ and $i=60^\circ$, the
  required overbrightness for a hot spot with $R_{\rm spot}=0.5M$
  would be 67\%.}
\begin{center}
\includegraphics{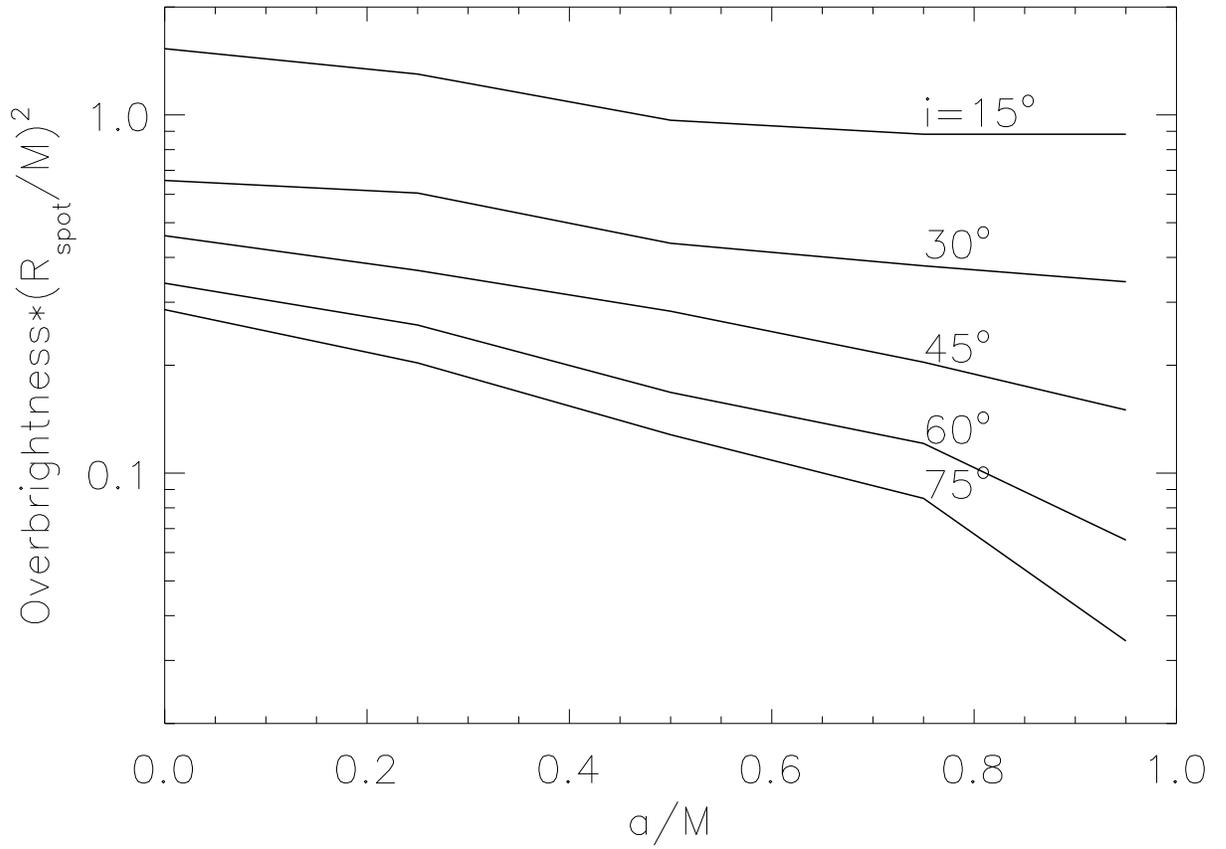}
\end{center}
\end{figure}

Figure \ref{plotsix} shows the required overbrightness of a flattened
Gaussian hot spot orbiting near the ISCO to produce a modulation with
rms amplitude of 1\% for a range of
inclinations and black hole spin parameters. In the limit of
a face-on accretion disk $(i=0^\circ)$, even an infinitely bright spot on
a circular orbit
will not produce a time-varying light curve. As the inclination
increases, the required overbrightness decreases, as the special
relativistic beaming focuses radiation toward the observer from a
smaller region of the disk, \textit{increasing} the relative
contribution from the hot spot. As the spin of the black hole
increases, the ISCO moves in toward the horizon and the velocity of a
trajectory near that radius increases, as does the gravitational
lensing, magnifying the contribution from the hot spot. 

Understandably, the required overbrightness is inversely proportional
to the area of the hot spot so $[\rm overbrightness]*R_{\rm spot}^2 =
\rm const$. For example, from Figure \ref{plotsix} we see that a black
hole binary with inclination $i = 60^\circ$ and spin $a/M=0.5$ would
require a spot size of $R_{\rm spot}= 0.5M$ with 67\%
overbrightness (e.g.\ 14\% temperature excess) to produce a 1\% rms
modulation in the light curve. This is well within the range of the
typical size and magnitude of fluctuations predicted
by 3D MHD calculations of 3-dimensional
accretion disks \citep{hawle01}. The hot spot model is
well-suited for simplified calculations of the X-ray emission from
these random fluctuations in the accretion disk. By adding the
emission from small, coherent hot spots to a steady-state
Shakura-Sunyaev disk, we can interpret the amplitudes and positions of
features in the QPO spectrum in terms of a model for the black hole
mass, spin, and inclination.

\begin{figure}[tp]
\caption{\label{plotseven} Frequency-integrated light curves of an
  orbiting hot spot at the ISCO of a Schwarzschild black hole for
  different disk inclination angles. The spot is moving in the
  $-\mathbf{e}_{\hat{\phi}}$ direction as in Figure \ref{plotfive}. For high
  inclination angles, the special relativistic beaming causes the light
  curve to become sharply peaked as the hot spot moves toward the
  observer.}
\begin{center}
\includegraphics{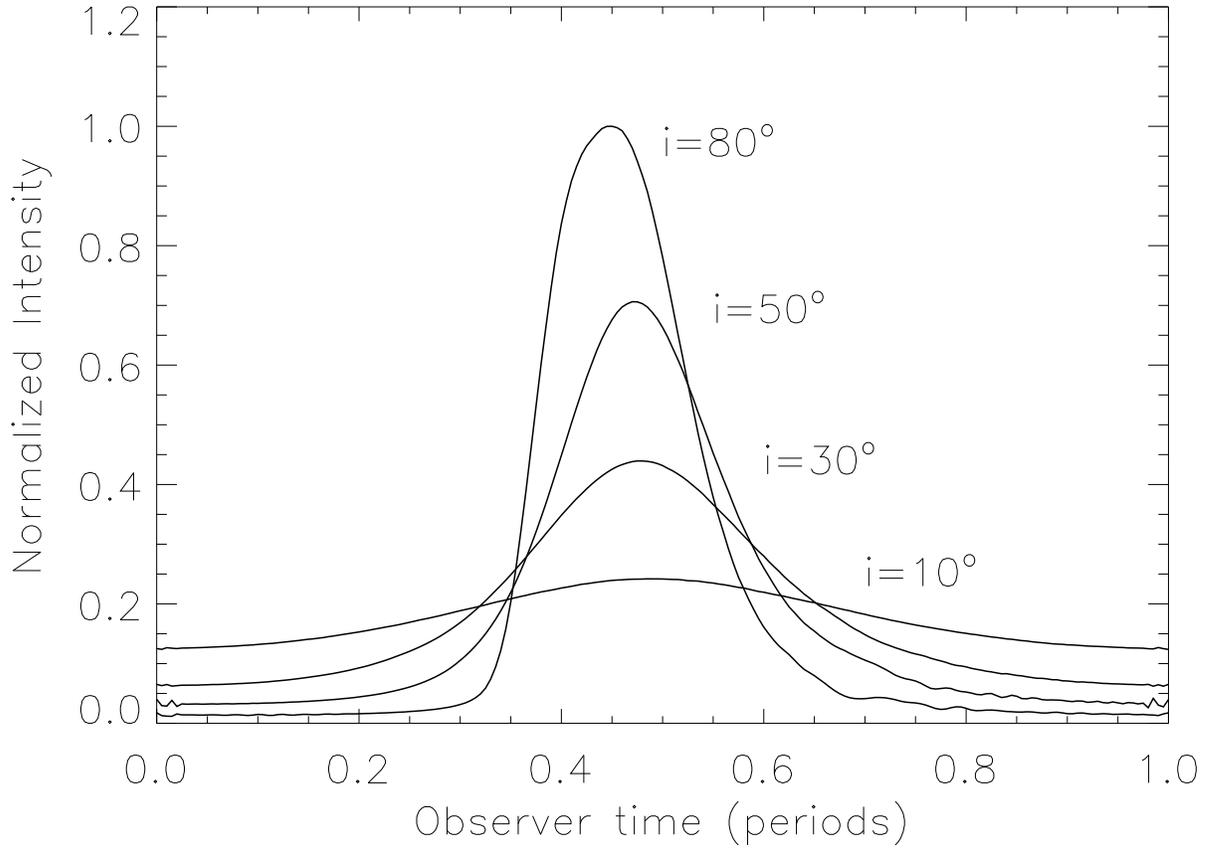}
\end{center}
\end{figure}

Considering the X-ray flux from the hot spot alone, the
frequency-integrated light curves for a variety of inclinations are 
shown in Figure \ref{plotseven}. All light curves are shown for one
period of a hot spot orbiting a Schwarzschild black hole at the
ISCO. As the inclination increases, the light curve goes from nearly
sinusoidal to being sharply peaked by special relativistic
beaming. Thus the shape of a QPO light curve may be used to
determine the disk inclination. With current observational
capabilities, it is not possible even for the brightest sources to get
a strong enough X-ray signal over individual periods as short as 3-5
ms to be able to differentiate between the light curves in Figure
\ref{plotseven}. Instead, the Fourier power spectrum can be used to
identify the harmonic features of a periodic or quasi-periodic light
curve over many orbits. Disks with higher 
inclinations will give more power in the higher harmonic frequencies,
due to the ``lighthouse'' effect, as the hot spot shoots a
high-power beam of photons toward the observer once per orbit,
approximating a periodic delta-function in time. 
 
\begin{figure}[tp]
\caption{\label{ploteight} (a) X-ray light curve of an
  orbiting hot spot with same parameters as in Figure
  \ref{plotfive}. (b) Fourier amplitude $a_n({\rm rms})$ of the
  above light curve with overbrightness of unity, normalized to
  the flux from a steady-state disk as in equation (\ref{arms}), showing the
  fundamental Kepler frequency at 220 Hz for $M=10M_\odot$. The
  non-sinusoidal shape of the light curve, due largely to beaming
  effects, is characterized by the declining power in the higher harmonic
  frequencies at $n220$ Hz, where $n > 1$ is an integer.}
\begin{center}
\scalebox{0.65}{\includegraphics{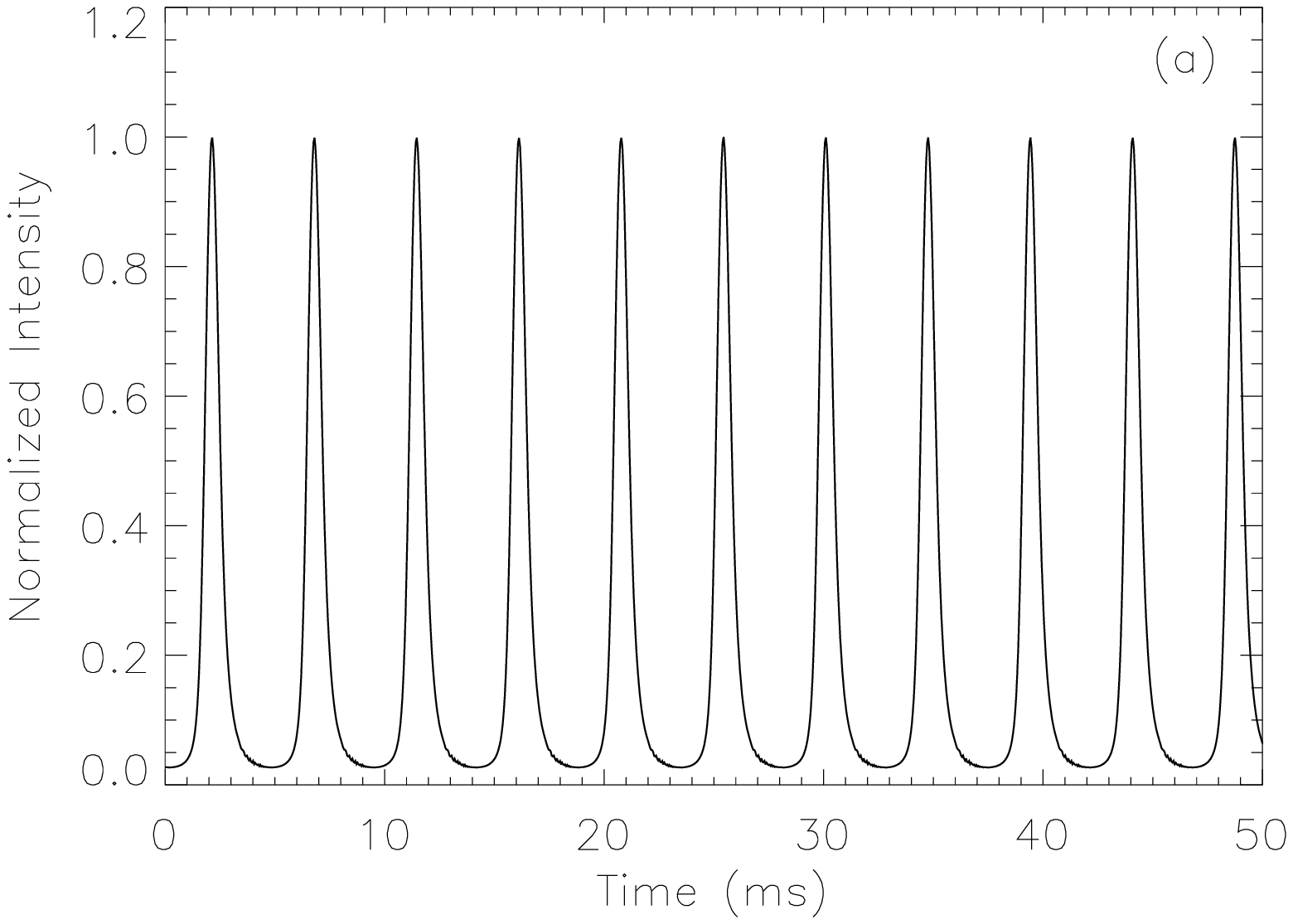}}
\scalebox{0.65}{\includegraphics{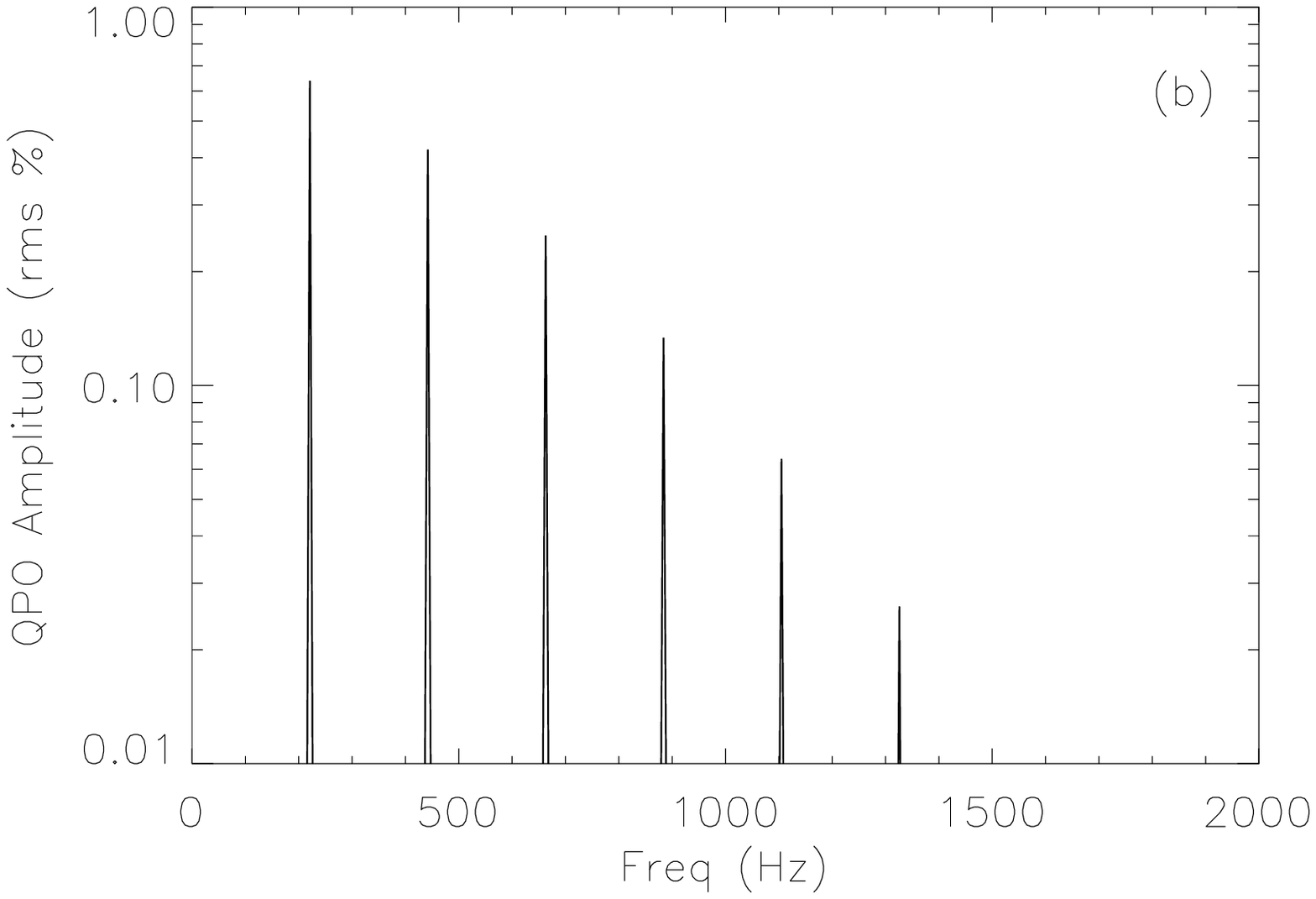}}
\end{center}
\end{figure}

Figure \ref{ploteight}a shows a sample section of such a light curve,
including only the X-ray flux from the hot spot, subtracting out the
steady-state flux from the disk. The sharp peaks in the light curve,
while unresolvable in the time domain, will give a characteristic
amount of power in the higher harmonics, shown in Figure
\ref{ploteight}b. Here we have normalized the rms amplitudes to
the background flux from the disk with a hot spot overbrightness of
100 \%. For a signal $I(t)$ with Fourier components $a_n$: 
\begin{equation}
I(t) = \sum_{n=0}^\infty a_n \cos(2\pi n t),
\end{equation}
we define the rms amplitude $a_n({\rm rms})$ in each mode $n>0$ as
\begin{equation}\label{arms}
a_n({\rm rms}) \equiv \frac{a_n}{\sqrt{2a_0}}.
\end{equation}
With this normalization, the rms defined in equation (\ref{rms}) can be
conveniently written 
\begin{equation}
{\rm rms} = \sqrt{\sum_{n>0} a_n^2({\rm rms})}.
\end{equation}

\begin{figure}[tp]
\caption{\label{plotnine} Fourier amplitude $a_n({\rm rms})$ in
  higher harmonic
  frequencies $\nu_n=n\nu_\phi$ as a function of orbital inclination
  to the observer, normalized as in equation (\ref{arms}). The hot spot
  has size $R_{\rm spot}=0.5M$, an overbrightness factor of 100\%, and
  is in a circular orbit at $R_{\rm ISCO}$ around a Schwarzschild 
  black hole.}
\begin{center}
\includegraphics{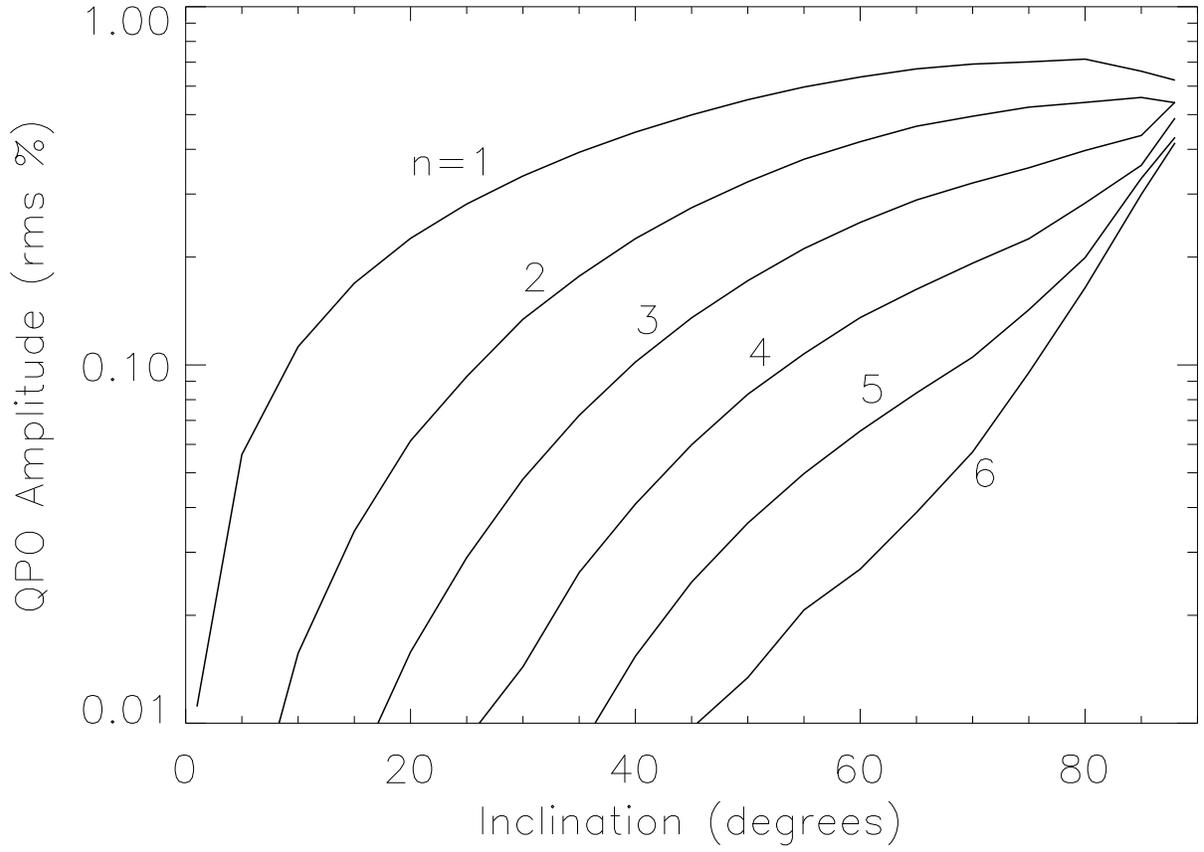}
\end{center}
\end{figure}

In Figure \ref{ploteight}b, the main peak at $f=220$ Hz corresponds to
the fundamental azimuthal frequency for
an orbit at the ISCO of a $10M_\odot$ Schwarzschild black
hole. In the limit where the light curve is a periodic
delta-function in time, there should be an equal amount of power
in all harmonic modes, because the Fourier transform of a periodic
delta-function is a periodic delta-function. However, even in the
limit of edge-on inclination $(i=90^\circ)$, unless the hot spot is
infinitessimally small and ultra-relativistic, the light curve will
always be a continuous function with some finite width
and non-zero minimum, thus contributing less and less power to the
higher harmonics. The harmonic dependence on inclination for a hot
spot orbiting a Schwarzschild black hole is shown in 
Figure \ref{plotnine}. Predictably, as the inclination increases,
we see that both the absolute and relative amplitudes of the higher
harmonics increase, almost to the limit of a periodic delta-function
when $i\to 90^\circ$. Interestingly, we find very little dependence of the
harmonic structure on hot spot size or shape, as long as the total
emission of the spot relative to the disk is constant (this emphasizes
the robustness of the simple hot spot model in interpreting an X-ray
power spectrum, without needing to include the detailed physics of the
disk perturbations). 

As the spin parameter increases for Kerr black holes, the ISCO moves
closer to the horizon, increasing the circular velocities of particles
on the ISCO and thus the Doppler shifts, giving broader
time-integrated spectra, as seen in Figure \ref{plotfour}. The phase
lag in time of the peak 
blueshift with respect to angular phase of the hot spot is
also amplified for these smaller values of $R_{\rm ISCO}$, giving light
curves that are asymmetric in time. The hot spot orbits around Kerr
black holes also experience more gravitational lensing as they move
through the strongly curved space-time near the horizon. 
This lensing becomes
more significant for larger inclinations as the black hole and the
emitting region approach collinearity with the observer, producing an
Einstein ring as $i\to 90^\circ$. This somewhat
counters the special relativistic beaming effect as the light curve
peaks over a broader phase of each period due to lensing. Thus the
harmonic power dependence on inclination for a Kerr light curve is in fact
smaller than that for the Schwarzschild geometry. However, in the event
that the accretion disk is actually rotating in a retrograde fashion
with respect to the black hole spin, the ISCO will move \textit{away}
from the horizon $[R_{\rm ISCO}(a/M=-0.99) = 8.97M]$, reversing the above
effects and giving a \textit{stronger} dependence on inclination.

\section{NON-CIRCULAR ORBITS}
One of the major unsolved puzzles motivating theoretical models of black
hole QPOs is the observation of multiple peaks in the high frequency
power spectrum \citep{lamb02}. As discussed above, any non-sinusoidal
light curve 
will contribute to Fourier power in harmonics at integer multiples of
the fundamental Kepler frequency. However, for at least two X-ray
binary systems (XTE J1550-564 and GRO J1655-40; possibly also GRS
1915+105), peaks are found with rational (but non-integer) frequency
ratios \citep{stroh01b,remil02}. In these
particular examples, significant power is measured around the
frequencies (184, 276 Hz) for XTE J1550-564 and (300, 450 Hz) for GRO
J1655-40, almost exactly a 2:3 commensurability in
frequencies, while GRS 1915+105 has peaks at 40 and 67 Hz. Following
the work of \citet{merlo99}, we investigate the 
possibility of these commensurabilities coming from integral
combinations of the radial and azimuthal coordinate frequencies of
nearly circular geodesics around a Kerr black hole. 

In a Newtonian point mass potential, the radial, azimuthal, and vertical
(latitudinal) frequencies $\nu_r,
\nu_\phi,$ and $\nu_\theta$ are identical, giving closed planar
elliptical orbits. For the Schwarzschild metric the vertical and
azimuthal frequencies are identical, giving planar rosette orbits that
are closed only for a discrete set of initial conditions. The Kerr
metric allows three unique coordinate frequencies, so geodesic orbits
in general can fill out a 
torus-shaped region around the black hole spin axis. When these
coordinate frequencies are rational multiples of each other, the
trajectories will close after a finite number of orbits. 

While there
is currently no clear physical explanation for why hot spots may
tend toward such trajectories, some recent evidence suggests the possible
existence of nonlinear resonances near geodesic orbits with integer
commensurabilities \citep{abram03}. Another important clue
may come from the fact that at these special orbits are closed,
perhaps enhancing the
stability of non-circular trajectories. The quasi-periodic nature of
the oscillations suggest the continual formation and subsequent
destruction of hot spots \textit{near}, but not exactly at, the
resonant orbits. For the purposes of this paper, we will take the
apparent preference for such orbits as given and concern ourselves
primarily with calculating the resulting light curves and power spectra. 

In geometrized units with $G=c=M=1$, coordinate time is measured in
units of $4.9\times 10^{-6} (M/M_\odot)$ sec. For example, an orbit
with angular frequency $\Omega_\phi=2\pi\nu_\phi=0.1$ around a $10M_\odot$
black hole would have an observed period of 3.1 ms, whereas the analogous
orbit around a supermassive black hole with mass $10^9M_\odot$ would
have a period of 8.6 hours. In these units, the
three fundamental coordinate frequencies for nearly
circular orbits are given by equation (\ref{omega_phi}) and
\citep{merlo99} 
\begin{equation}\label{omega_theta}
\Omega_\theta = 2\pi\nu_\theta = \Omega_\phi \left[1\mp
\frac{4a}{r^{3/2}}+\frac{3a^2}{r^2}\right]^{1/2}
\end{equation}
and
\begin{equation}\label{omega_r}
\Omega_r = 2\pi\nu_r =
\left[\frac{r^2-6r\pm8ar^{1/2}-3a^2}{r^2(r^{3/2}\pm a)^2}\right]^{1/2},
\end{equation}
where the upper sign is taken for prograde orbits and the lower sign
is taken for retrograde orbits. The radial frequency approaches zero
at $r\to R_{\rm ISCO}$, where geodesics can orbit the black hole many
times with steadily decreasing $r$. In the limit of zero spin and large
$r$, the coordinate frequencies reduce to the degenerate
Keplerian case with $\Omega_\phi = \Omega_\theta = \Omega_r = r^{-3/2}$.

\begin{figure}[tp]
\caption{\label{plotten} Radius of prograde orbits with
  commensurate frequencies $\nu_r:\nu_\phi = $(1:3, 1:2, 2:3) (solid
  lines) as a function of dimensionless spin parameter $a/M$. The ISCO
  (dashed line) corresponds to $\nu_r:\nu_\phi = 1:\infty$. Also
  shown are the respective orbital frequencies $\nu_\phi$
  at these radii for a black hole with mass $10M_\odot$ (dot-dashed
  lines).
}
\begin{center}
\includegraphics{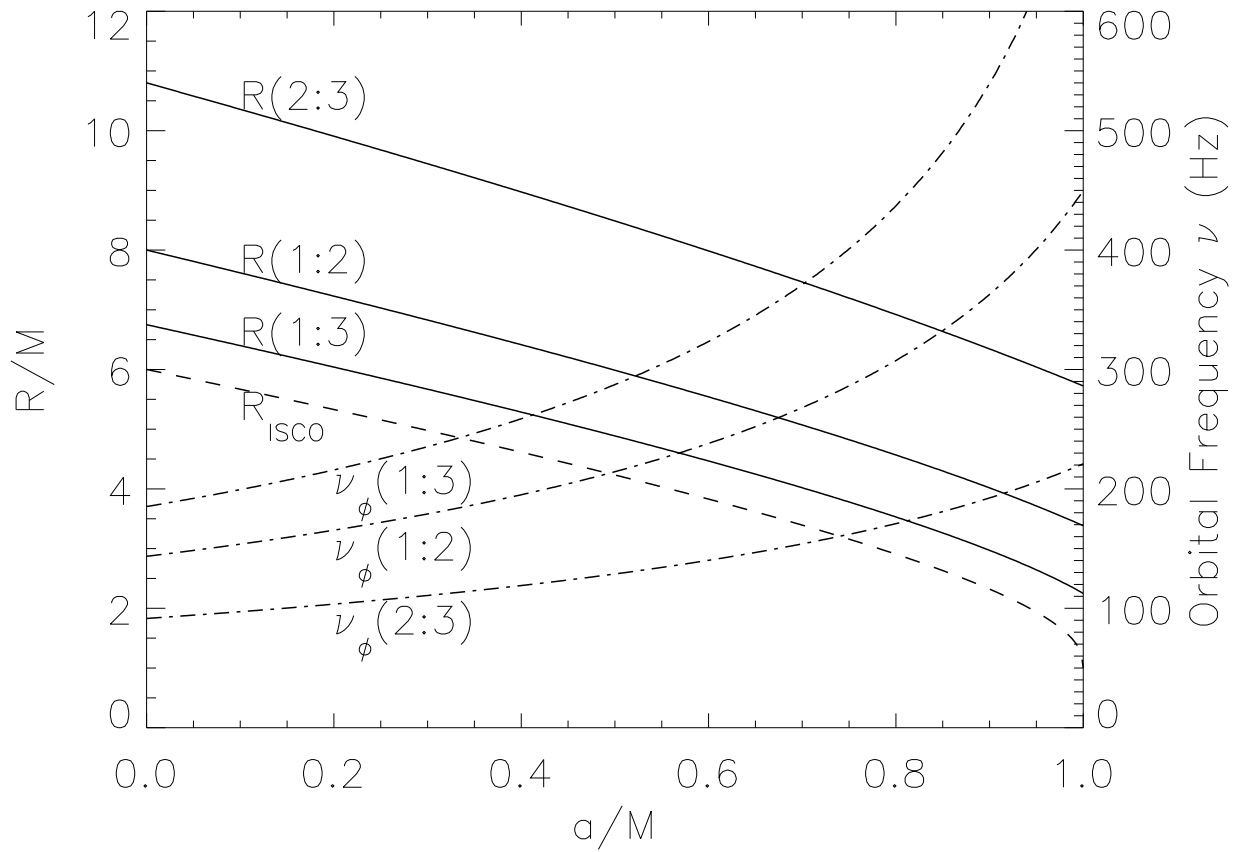}
\end{center}
\end{figure}

To model the 2:3 frequency commensurability, we begin by looking for
perturbed circular planar orbits where the radial frequency $\nu_r$ is
one-third the azimuthal frequency $\nu_\phi$. Since the orbits are
nearly circular, the fundamental mode of the light curve should peak
at the azimuthal frequency with additional power in beat modes at
$\nu_\phi \pm \nu_r$. For $\nu_r$:$\nu_\phi$ = 1:3, the power spectrum
should have a triplet of peaks with frequency ratios 2:3:4. These
commensurate orbits can be found easily from equations (\ref{omega_phi})
and (\ref{omega_r}) and solving for $r$:
\begin{equation}
r^2 -6r\pm8ar^{1/2}-3a^2=\pm\frac{r^2}{9}(r^{3/2}\pm a).
\end{equation}
Figure \ref{plotten} shows the radius (solid lines) of these special
orbits as a function of spin parameter. Also shown (dashed line) is
the inner-most stable circular orbit for prograde trajectories. The
position of the 1:3 commensurate orbits follows 
closely outside the ISCO, suggesting a connection
between the high frequency QPOs and the black hole ISCO. However,
other integer commensurabilities such as 1:2, 2:5, or 1:4 also closely
follow the ISCO curves for varying $a$, so the proximity to the ISCO
alone is probably not enough to explain the hot spot preference for
these specific coordinate frequencies. It is important to note that
any given black hole source will have a constant value of $a$,
certainly over the lifetime of our observations. Thus, we may need to
observe many more sources like XTE J1550-564 and GRO J1655-40 in order
to better 
sample the parameter space of Figure \ref{plotten} and thus the
connection between certain preferred orbits and the black hole
ISCO. 

\begin{figure}[tp]
\caption{\label{ploteleven} (a) X-ray light curve of
  a hot spot orbit with $\nu_\phi=3\nu_r$, $e=0.089$, $M=10M_\odot$,
  $a/M=0.5$, $i=60^\circ$, and $R_{\rm spot}=0.5M$. (b) The Fourier
  amplitude $a_n({\rm rms})$ of the above light curve, normalized as
  in Figure \ref{ploteight}b, showing the fundamental Kepler frequency at
  $\nu_\phi=285$ Hz and beat modes at $\nu = n\nu_\phi\pm \nu_r$.}
\begin{center}
\scalebox{0.65}{\includegraphics{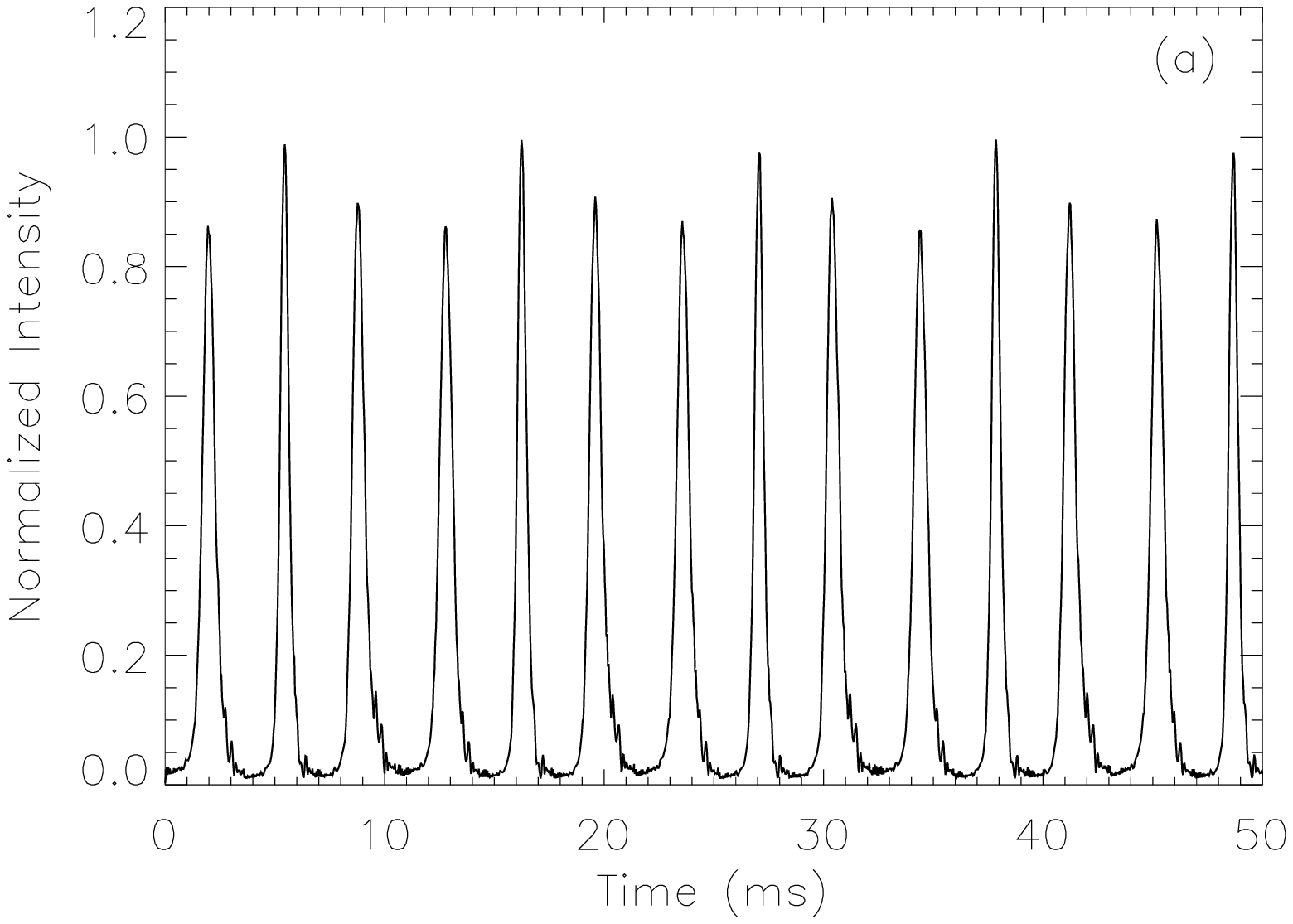}}
\scalebox{0.65}{\includegraphics{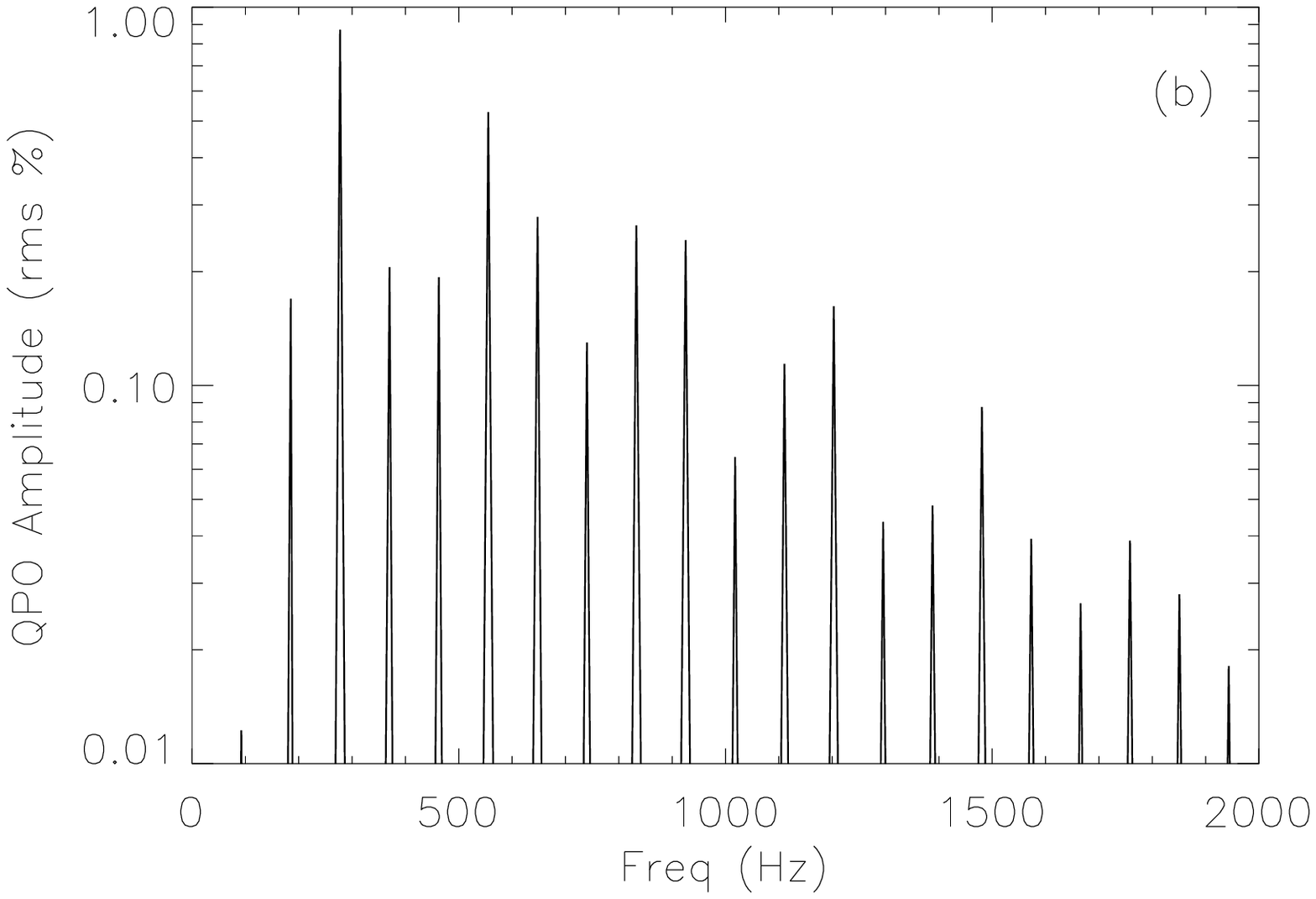}}
\end{center}
\end{figure}

A 1:3 commensurate trajectory moves through three revolutions in
azimuth for each radial period,
forming a closed rosette of three ``layers.'' For such rosettes, the
eccentricity can be defined as 
\begin{equation}
e \equiv \frac{r_{max}-r_{min}}{r_{max}+r_{min}}.
\end{equation}
The time-dependent light curve for an prograde orbit with eccentricity
$e=0.089$, spin $a/M=0.5$, and inclination $i=60^\circ$ is shown in Figure 
\ref{ploteleven}a. The time axis begins at the point when the hot
spot is at apocenter, moving away from the observer. Thus the first
and third peaks come from the hot spot moving toward the observer at
a relatively larger radius, while 
the second, higher peak is caused by the emitter moving toward the
observer through pericenter at a higher velocity, giving a larger
blue-shift and thus
beaming factor. The combined Doppler beaming and gravitational lensing
causes the peak following the pericenter peak to be slightly larger,
as the emitter is moving away from the black hole yet the light is
focused more toward the observer.

The power spectrum for this light curve is shown in Figure
\ref{ploteleven}b, with the strongest peaks at the azimuthal
frequency of $\nu_\phi = 285$ Hz and its first harmonic at $2\nu_\phi =
570$ Hz for $M=10M_\odot$.  Even for this modest deviation from
circularity, there is significant power in the frequencies
$\nu_\phi \pm \nu_r$. The beating of the fundamental $\nu_\phi$ with
the radial frequency $\nu_r = (1/3)\nu_\phi= 95$ Hz gives the
set of secondary peaks at $(2/3)\nu_\phi$ and
$(4/3)\nu_\phi$. Additional peaks 
occur at beats of the harmonic frequencies $n\nu_\phi \pm \nu_r$. It
is interesting to note that there is not significant power in the
radial mode at $\nu = 95$ Hz, but only in the beats with the
fundamental azimuthal frequency and its harmonics. However, in the
limit of a face-on orientation $(i\to 0^\circ)$, the radial frequency should
dominate the light curve variation as the gravitational and transverse
Doppler red-shift modulate the intensity as a function of the hot
spot's radial coordinate. The radial mode should also be present in
the limit of an edge-on orientation $(i\to 90^\circ)$, as gravitational
lensing becomes more important, and the hot spot will experience more
magnification when closer to the black hole.

\begin{figure}[tp]
\caption{\label{plottwelve} Power in low-order harmonics and beat
  modes with frequencies $\nu = n\nu_\phi \pm \nu_r$, as a function of
  disk inclination angle. The hot spot trajectory is the same as in
  Figure \ref{ploteleven}, with $a_n({\rm rms})$ normalized as in
  Figure \ref{plotnine}.
  The curves are labeled by the ratio $\nu/\nu_\phi$.}
\begin{center}
\includegraphics{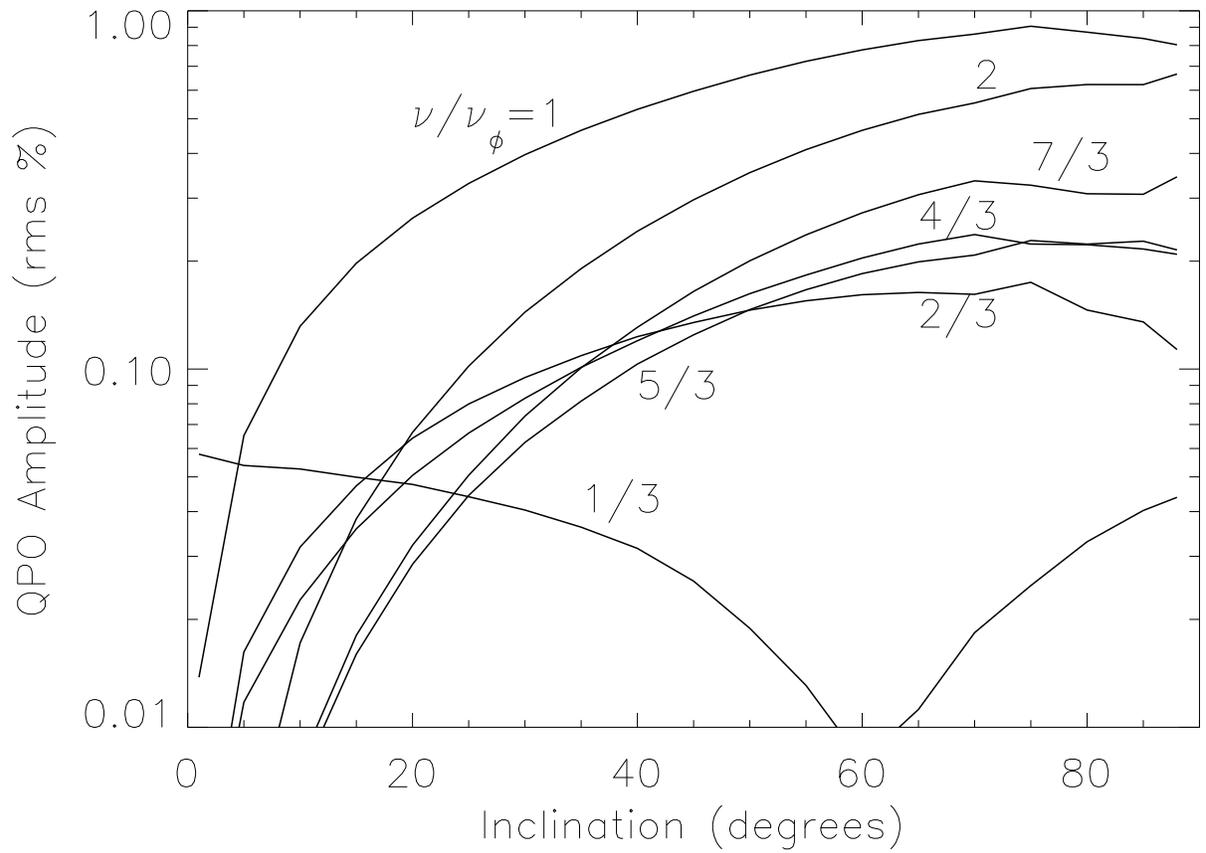}
\end{center}
\end{figure}

Figure \ref{plottwelve} shows the dependence on
inclination of the lower order harmonic and beat modes. At low
inclinations, the radial frequency $(\nu/\nu_\phi =1/3)$
contributes significant
power, while at higher inclinations, the first harmonic of the
azimuthal mode begin to dominate with similar behavior
to the circular orbits shown in Figure \ref{plotnine}. Along with
increasing power at $2\nu_\phi$, there is also increasing power in the
radial beats of the first harmonic at $2\nu_\phi \pm \nu_r =
(5/3)\nu_\phi, (7/3)\nu_\phi$. For a $10M_\odot$ black
hole, all these frequencies should be observable within the timing
sensitivity of RXTE.

To further explore the constraints of our model, we
investigated the effect of orbital eccentricity on the QPO
power. Maintaining a 1:3 commensurability between radial and
azimuthal frequencies, we calculated the light curves for a range of
eccentricities $0\le e \le 0.2$. As expected, the beat modes at $\nu=\nu_\phi
\pm \nu_r$ have more power for more eccentric orbits, as the radial
variation of the emitter becomes larger. At the same time, the first
harmonic at $\nu=2\nu_\phi$ provides relatively less power with
increasing eccentricity. This is best understood as the ``picket
fence'' character of the light curve becomes modulated in amplitude
and frequency from peak to peak, i.e. for each 3-peak cycle, the time
between peaks 1-2, 2-3, and 3-1 are not all the same, damping the
harmonic overtone. Interestingly, the Fourier power in the beat modes
appears to saturate at a moderate eccentricity of $e\approx 0.1$, suggesting
highly eccentric geodesics are not necessary or even helpful for
producing power in the beat frequencies.

While there is significant evidence for higher frequency harmonic
and beat modes in the QPO power spectrum of XTE J1550-564, the Fourier
power is clearly dominated by the two frequencies $184$ and $276$ Hz
\citep{remil02},
corresponding to $\nu_\phi-\nu_r$ and $\nu_\phi$ in our model. What
are the physical mechanisms that could damp out the higher
frequency modes? One possible explanation is in the geometry of the
hot spot. As explained in Section 4, in order to produce the
power observed in QPOs, the relative flux coming from the hot
spot must be some finite fraction of that of the disk (typically
$10^{-3}-10^{-2}$ for a QPO amplitude of $1-5\%$), so the hot
spot must have some minimum size or it would not produce enough
emission to be detected above the background disk. Yet if the
hot spot is too large, it would be sheared by differential rotation in
the accretion disk and not survive long enough to create the coherent
X-ray oscillations that are observed. As mentioned above,
we find that the relative QPO power in different modes is not
sensitive to the size of the hot spot 
$R_{\rm spot}$. Three-dimensional MHD simulations \citep{hawle01} show
a range of density and temperature fluctuations consistent with
the hot spot size and overbrightness factor predicted by our model
in conjunction with the observations.

\begin{figure}[tp]
\caption{\label{plotthirteen} Power spectrum for a hot spot with
  same trajectory as in Figure \ref{ploteleven}, with the emission
  region sheared along the geodesic into an arc of length (a) $180^\circ$
  and (b) $360^\circ$. For the shorter arc (a), the power is still peaked
  at the fundamental frequency $\nu_\phi=285$ Hz, while the extended
  arc (b) produces more power in the beat frequency $\nu_\phi-\nu_r =
  190$ Hz.}
\begin{center}
\scalebox{0.65}{\includegraphics{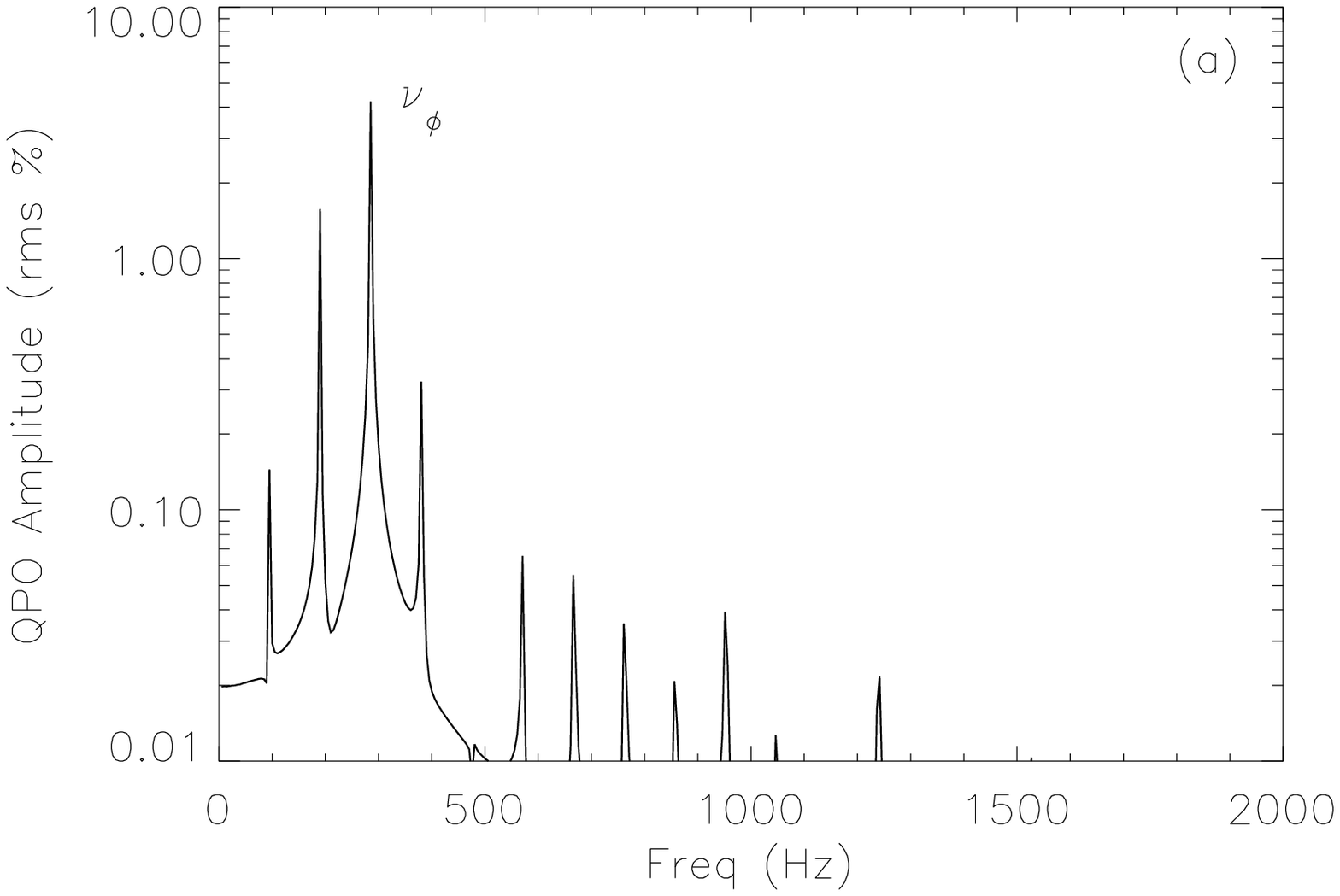}}
\scalebox{0.65}{\includegraphics{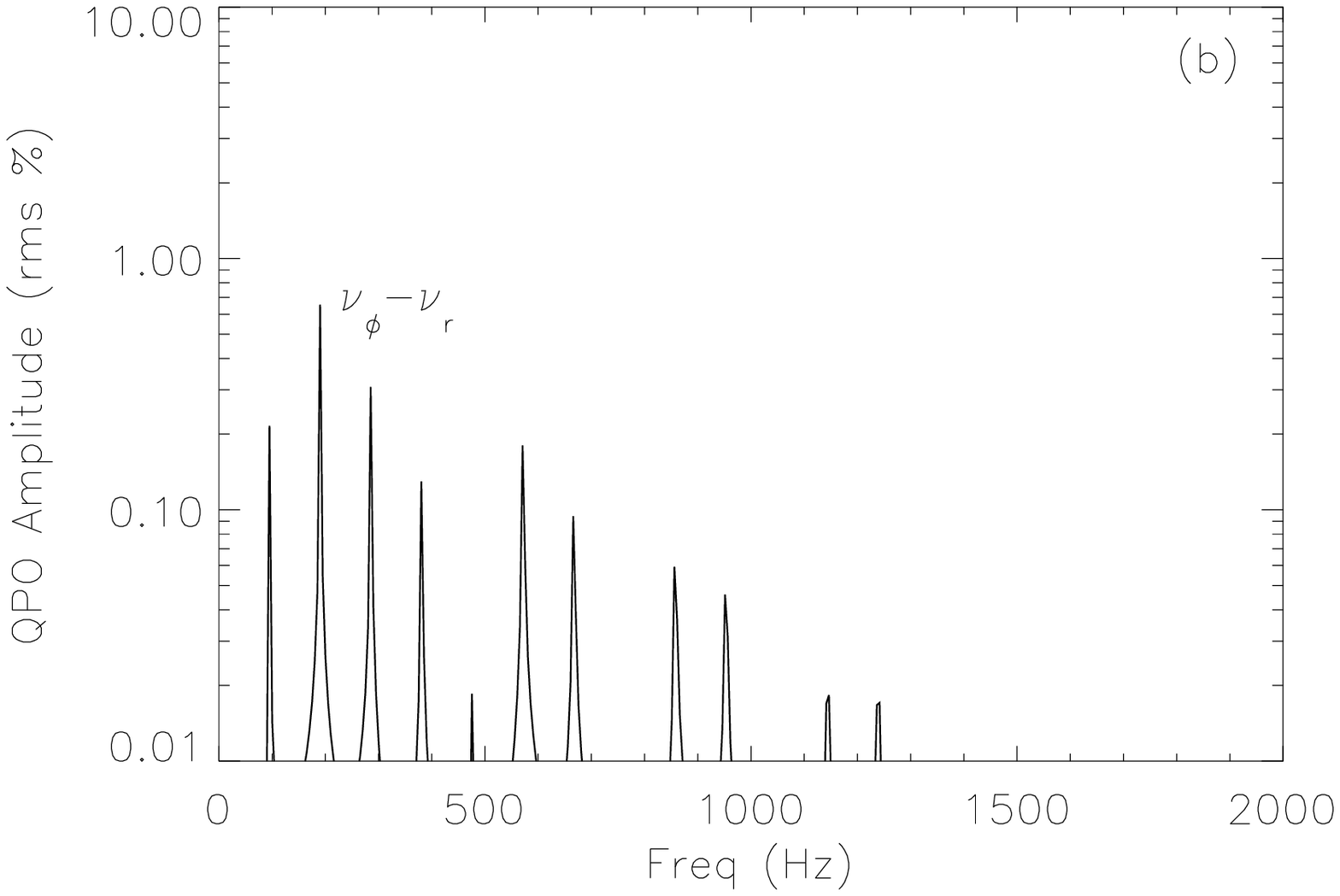}}
\end{center}
\end{figure}

It also appears from simulations that as the hot spot is formed in the disk,
differential rotation will tend to shear a finite region of gas as
it follows a geodesic orbit around the black hole, modifying the shape
of the hot spot into an arc in azimuth. In the limit that
the emission region could be sheared into a ring of arc length $\Delta
\phi = 360^\circ$, the fundamental mode and its harmonics would be
essentially removed, leaving power only in the radial
modulation. Indeed, as shown in Figure
\ref{plotthirteen}a, for an arc length of $\Delta\phi=180^\circ$, the higher
frequency modes at $\nu = 2\nu_\phi$ and $\nu=2\nu_\phi \pm \nu_r$ are
strongly suppressed, while still maintaining a significant amount of
power in the fundamental beat modes $\nu_\phi\pm\nu_r$. The total QPO
power also increases as the area of the emission region increases
relative to the circular hot spot geometry. 

However, when we
allow the arc to be sheared into a ring with $\Delta\phi=360^\circ$, the
total QPO power is actually decreased as the differential beaming is
essentially eliminated by the extended emission region: there
is always some portion of the arc moving toward the
observer. The resulting modulation is then more weighted to
the first radial beat mode at $\nu_\phi-\nu_r$, as seen in Figure
\ref{plotthirteen}b. It is not intuitively obvious why the
$\nu_\phi-\nu_r$ mode is dominant while the $\nu_\phi+\nu_r$ mode
($\nu$ = 368 Hz) is much weaker in the arc geometry. If anything, this
is a strong argument for the necessity of a full ray-tracing calculation
of the hot spot light curves when predicting QPO power spectra, as it
clearly gives information unavailable to simple analysis of the
geodesic coordinate frequencies.

This behavior offers a plausible explanation for the two major types
of QPOs described in \citet{remil02}. Type A outbursts have
more total power in the HFQPOs, with a major peak at 276 Hz and a
minor peak at 184 Hz. Type B outbursts have most of the QPO power
around 184 Hz and a smaller peak around 276 Hz and less overall
power in the high frequency region of the spectrum. Thus we propose
that Type A outbursts are coming from more localized
hot spot regions, while Type B outbursts come from a more extended ring
geometry.  

In addition to the commensurate high-frequency QPOs observed in
sources like XTE J1550-564, there are also strong low-frequency QPOs
observed at the same time with frequencies in the range
$5-15$Hz. There have been suggestions that these low-frequency QPOs
may be caused by the Lens-Thirring precession of the disk near the
ISCO, also known as ``frame dragging'' \citep{marko98, merlo99, abram01,
  remil02}. For geodesic orbits out of the plane perpendicular to the
black hole
spin, the latitudinal frequency $\Omega_\theta$ of massive particles
is not equal to the azimuthal frequency $\Omega_\phi$ [see eqs.\
(\ref{omega_theta}) and (\ref{omega_phi})], leading to a
precession of the orbital plane with frequency
\begin{equation}
\Omega_{LT} \equiv |\Omega_\theta - \Omega_\phi|.
\end{equation}
For black holes with the mass and spin used
above $(M = 10M_\odot, a/M = 0.5)$, the frame-dragging
frequency, as calculated at the radius corresponding to the
commensurability $\nu_r$:$\nu_\phi$=1:3, is somewhat higher than that
observed in the low-frequency QPOs from XTE J1550-564. The Type A QPO
peaks at 12 and 276 Hz appear to be consistent with a black hole
mass of $8.9 M_\odot$ and spin parameter of $a/M = 0.35$, quite
similar to the values used throughout much of this paper. For the BH
binary GRO J1655-40, we can fit the QPOs at 18 and 450Hz with a mass
of $5.1 M_\odot$ and spin $a/M=0.28$, slightly less than the published
mass range of $5.5-7.9 M_\odot$ \citep{shahb99}. If we relax 
the requirement of matching the LFQPOs and only fit the HFQPOs with a 1:3
coordinate frequency commensurability, there remains
a 1-dimensional degeneracy in the mass-spin parameter space. Based
solely on the HFQPOs, for XTE
J1550-564 with $8.5<M/M_\odot<11.5$, the range of spin parameters
would be $0.33<a/M<0.6$ and for GRO J1655-40, the spin would be in the
range $0.35<a/M<0.66$. These results are shown in Table
\ref{tableone}. 

To get a more quantitative feeling for the effect of Lens-Thirring
precession on the power spectrum, we investigated hot spot
orbits with initial trajectories inclined to the plane of the disk:
$v^\theta \ne 0$. This is much like changing the observer's
inclination with a period of $2\pi/\Omega_{LT}$. Thus we see
additional modulation in the hot spot light curve at the
``double-beat'' modes $\nu_\phi \pm \nu_r \pm \nu_{LT}$. We find that,
for modestly inclined hot spot orbits $(i_0 = \pm 5^\circ)$,
the contribution to the power spectrum at Lens-Thirring frequencies is
quite small ($<1\%$ of total power) for the basic circular hot spot
geometry. This relative contribution increases with arc 
length as the spot becomes a ring precessing about the spin axis,
consistent with the relative power in LFQPOs and HFQPOs
in the Type A (more high frequency power than low frequency) and Type
B (more low frequency power) sources described
above. 

Under the premise that the HFQPO commensurate frequencies are
caused by the geodesic motion of a sheared, overbright region in the
disk, in Table \ref{tabletwo} we show the best fit parameters for the
Type A and Type B QPO
spectra from XTE J1550-564 [cf. Table 1 in \citet{remil02}]. Guided
also by the (somewhat speculative) assumption that the LFQPOs come
from the Lens-Thirring precession of the hot spot orbital plane, we
predict a black hole mass and spin. Using a fixed inclination of
$70^\circ$, we can match the frequencies and amplitudes of the observed
HFQPO peaks (and, equally important, the lack of power at certain
frequencies) for both Type A and Type B outbursts. Setting constant the
eccentricity $e=0.1$, geodesic inclination to the disk $i_0=5^\circ$, and
the overbrightness to be a factor of unity, we fit the hot spot radius
and arc length to match the observations. Being able to match the QPO
rms amplitudes of the 
peaks (or lack thereof) at 92, 184, 276, and 368 Hz, for at least two
different types of X-ray outburst, shows the robustness of our simple
model in explaining these phenomena.

However, even with the sheared arc emission, many observations
still show significantly more power in the LFQPOs than can be
explained solely from the Lens-Thirring precession of a geodesic near
the ISCO. Coupled with the difficulty in simultaneously fitting the
mass and spin to three coordinate frequencies in a manner consistent
with spectroscopic mass predictions, it seems likely that the LFQPOs
may be caused by some other mechanism in the disk that is related only
indirectly to the high-frequency hot spot emission. Another possibilty
is that the thin, warped disk model breaks down near the ISCO,
allowing more complicated emission geometries and thus amplifying the
effects of Lens-Thirring precession.

\section{CONCLUSIONS}
We have developed a hot spot model to help understand the high-frequency
QPOs observed in accreting black holes. While the model does not
consider radiation or hydrodynamic pressures, it is fully relativistic
and calculates exact geodesics for both massive particles and
photons. Treating the hot spot as a monochromatic, isotropic emitter, we
have calculated the time-dependent spectra of X-ray lines produced as
the hot spot orbits the black hole. 

The model makes a number of general predictions of the Fourier power
of the X-ray light curve as a function of inclination and black hole
spin, and is also able to explain QPO observations from the black hole
binaries XTE J1550-564 and GRO J1655-40. Simply by matching the
locations of the low-frequency and high-frequency QPOs with the
coordinate frequencies (under the condition $\nu_\phi=3\nu_r$), we can
determine the black hole mass and spin. Relaxing the LFQPO constraint,
the spin can still be determined uniquely for a given mass, which in turn
can often be measured independently with the inclination and radial
velocity of the companion star. 

By matching the amplitudes of various peaks observed in XTE J1550-564,
we have explored the model parameters such as the hot spot
size, shape, and the overbrightness relative
to a steady-state background disk. The predicted magnitude of these
fluctuations are within the range predicted by 3D MHD calculations of the
accretion disk. Future work will investigate the effect of multiple
hot spots of various size, emissivity, and lifetime, as guided by MHD
calculations. Observations of additional sources
with commensurate frequency QPOs may help us further constrain the
hot spot model and better understand the connection between the
LFQPOs and HFQPOs. 

Some of the physical problems with the hot spot model have
been raised by \citet{marko00}, as discussed above.
Many of these points are addressed in our model. First, unlike
\citet{stell98,stell99}, we only
attempt to explain a set of QPO data from \textit{black hole} binaries,
which differ qualitatively from neutron star binaries in many ways,
e.g.\ lacking strong global magnetic fields and thermonuclear
activity. And perhaps most significant, black
holes have no rotating surface to interact with the accreting matter
and provide additional confusion to the QPO power spectrum.
Our model produces light curves with power spectra consistent with
black hole observations even with low eccentricity hot spot
orbits. 

Because they do not include ray tracing in their calculations,
\citet{marko00} are unable to model many relativistic effects,
including the gravitational lensing of
the hot spot source, which can be quite significant for systems with
moderate to high inclination angles $(i \ge 60^\circ)$. Since we calculate
the actual X-ray modulation from the orbiting hot spots, we predict
both the location and amplitude of every peak in the light curve power
spectrum, which cannot be done by analyzing the BH coordinate
frequencies alone. By introducing a perturbation on circular orbits
near the ISCO, additional peaks begin to appear in the power
spectrum, caused by beats of the azimuthal and radial frequencies
$\nu_\phi$ and $\nu_r$. The dependence of the relative power in the
different peaks on inclination and spin helps to constrain the
details of the hot spot model in explaining the HFQPOs, particularly the
2:3 commensurability observed in the power spectra from XTE
J1550-564 and GRO J1655-40. 

As an additional parameter, we introduce
a finite arc length for the emission region, motivated by the
shearing of the hot spot by differential rotation in the disk. The
spreading of the hot spot in azimuth leads to suppression of the
higher QPO modes, in agreement with observations.
We have also examined the possibility of Lens-Thirring precession
for non-planar orbits as an explanation for the low-frequency QPOs
that have been observed coincident with the HFQPOs yet often with even
stronger Fourier power. The predicted power spectra from non-planar
precessing arcs are consistent with observations of XTE J1550-564 if we
associate Type A outbursts with hot spot arcs of $\Delta \phi \approx
180^\circ$ and Type B outbursts with hot spot rings of $\Delta \phi
\approx 360^\circ$. However, the difficulty in matching the LFQPOs
amplitude and frequency with a single hot spot geodesic suggests the
low-frequency modulations may be caused by a different mechanism or
perhaps our disk geometry is too simplistic.

One major remaining issue with the hot spot model is the preferred
location of the geodesic that gives rise to 1:3 coordinate
frequencies. Why should the orbital frequencies favor integer ratios,
and why should the preferred ratio be 1:3 and not 1:2 or 1:4? It is
possible that detailed radiation-hydrodynamic 
calculations with full general relativity will be required to answer this
question. Perhaps the non-circular orbits can only survive along
closed orbits such as these to somehow avoid destructive
intersections. Or there may be magnetic interactions with the black
hole itself, analogous to the Blandford-Znajek process, that lock the
accreting gas into certain preferred
trajectories \citep{wang03}. For now, we are forced to leave this as
an open question unanswered by the geodesic hot spot model.

A less difficult, yet still unanswered problem is the explanation of
the widths of the QPO peaks. As it stands, our hot spot model predicts
purely periodic light curves and thus power spectra made up of
delta functions. If there \textit{is} some physical mechanism 
that preferentially focuses accreting material onto eccentric orbits
at specific radii, then it is likely that these hot spots are forming
and then being destroyed as a continual processes. The superposition
of many hot spots around the same orbit, all with slightly different initial
trajectories, could explain the quasi-periodic nature of the power
spectrum: the phase decoherence of the hot spots would cause a natural
broadening of the strictly periodic signal from a single spot.
With the computational framework in place, this question can be
answered by modeling a whole collection of hot spots and arcs
continually forming and evolving in shape and emissivity. The
particular physical parameters for these hot spots can be derived from
published MHD calculations such as \citet{hawle01}. With the basic
ray-tracing and radiation transport methods in place, it should be
possible to use our code as a ``post-processor'' to analyze the
results of these 3D simulations to simulate X-ray light curves and
spectra from a realistic accretion disk.

A final piece of the black hole QPO puzzle is the spectral behavior of
the source during outburst. \citet{remil02} have shown there is a
large range of X-ray fluxes with different relative contributions from
a power-law component and a disk bolometric component. These relative
and absolute fluxes seem to be correlated with the amount of power in
both the LFQPOs and the HFQPOs. Future work on 3-dimensional disks
and more detailed radiative transfer models should give us important
insights into understanding this spectral behavior and its relation to
the QPO power.

\vspace{1cm}
We thank Ron Remillard and Mike Muno for helpful discussion and
comments on the manuscript. This work was supported by NASA grant
NAG5-13306.

\newpage

\begin{table}[tp]
\caption{\label{tableone} Black hole parameters for hot spot
  model, matching low-frequency and high-frequency QPOs to geodesic
  coordinate frequencies}
\begin{center}
\begin{tabular}{llcc}
  Black Hole Parameters &  & XTE J1550-564 & GRO J1655-40 \\
  \hline
  & BH Mass & $8.9M_\odot$ & $5.1M_\odot$ \\
  & BH Spin & $0.35M$ & $0.28M$ \\
  & $R_{\rm ISCO}$ & $4.8M$ & $5.05M$ \\
  & Inclination & $70^\circ$ & $65^\circ$ \\
  \hline\hline
  Geodesic Frequencies & & \\
  \hline
  & $r_0$ & $5.54M$ & $5.77M$ \\
  & $\nu_{\rm LT}$ & 12 Hz & 17 Hz \\
  & $\nu_r$ & 92 Hz & 150 Hz \\
  & $\nu_\phi$ & 276 Hz & 450 Hz 
\end{tabular}
\end{center}
\end{table}

\begin{table}[tp]
\caption{\label{tabletwo} QPO Amplitudes of hot spot/arc model for XTE
  J1550-564}
\begin{center}
\begin{tabular}{lccc}
  Parameter & & Type A & Type B \\
  \hline
  $R_{\rm spot}$ & & $0.3M$ & $0.5M$ \\
  arc length & & $200^\circ$ & $320^\circ$ \\
  eccentricity & & $0.1$ & $0.1$ \\
  inclination to disk & & $5^\circ$ & $5^\circ$ \\
  overbrightness & & 100\% & 100\% \\
  \hline
  Amplitude (mode) & Frequency (Hz) & rms(\%) & rms(\%) \\
  \hline
  $a(\nu_{\rm LT})$ & 12 & 0.63 & 2.1 \\
  $a(\nu_r)$ & 92 & 0.48 & 0.89 \\
  $a(\nu_\phi-\nu_r)$ & 184 & 1.3 & 2.2 \\
  $a(\nu_\phi)$ & 276 & 3.2 & 0.42 \\
  $a(\nu_\phi+\nu_r)$ & 368 & 0.20 & 0.23 \\
\end{tabular}
\end{center}
\end{table}

\newpage

\end{document}